\documentclass[manuscript,nonacm,authorversion,screen]{acmart}
\usepackage{caption}
\usepackage{graphicx}
\usepackage[bb=boondox,bbscaled=.95,cal=boondoxo]{mathalfa} 

\AtBeginDocument{%
  \providecommand\BibTeX{{%
    \normalfont B\kern-0.5em{\scshape i\kern-0.25em b}\kern-0.8em\TeX}}}

\setcopyright{acmcopyright}
\copyrightyear{2023}
\acmYear{2023}
\acmDOI{}

\acmConference[CSCW 2025]{CSCW 2025: ACM SIGCHI Conference on Computer-Supported Cooperative Work & Social Computing}{CSCW, 2025}{TBD}
\acmISBN{}
\begin{document}

\title[My Views Do Not Reflect Those of My Employer]{My Views Do Not Reflect Those of My Employer: Differences in Behavior of Organizations' Official and Personal Social Media Accounts}

\author{Esa Palosaari}
\email{esa.palosaari@aalto.fi}
\orcid{0000-0003-4880-5548}
\affiliation{%
  \institution{Aalto University}
  \streetaddress{Konemiehentie 2, 02150}
  \city{Espoo}
  \country{Finland}
  \postcode{43017-6221}
}

\author{Ted Hsuan Yun Chen}
\orcid{0000-0002-3279-8710}
\email{ted.hsuanyun.chen@gmail.com}
\affiliation{%
  \institution{George Mason University}
  \streetaddress{4400 University Drive}
  \city{Fairfax}
  \state{Virginia}
  \country{United States of America}
  \postcode{22030}
}

\author{Arttu Malkamäki}
\email{arttu.malkamaki@helsinki.fi}
\orcid{0000-0002-7751-8831}
\affiliation{%
  \institution{University of Helsinki}
  \streetaddress{}
  \city{Helsinki}
  \country{Finland}}
\email{}

\author{Mikko Kivelä}
\email{mikko.kivela@aalto.fi}
\orcid{0000-0003-2049-1954}
\affiliation{%
  \institution{Aalto University}
  \streetaddress{Konemiehentie 2, 02150}
  \city{Espoo}
  \country{Finland}
  \postcode{43017-6221}
}

\renewcommand{\shortauthors}{Palosaari, Chen, Malkamäki, \& Kivelä}

\begin{abstract}
  On social media, the boundaries between people's private and public lives often blur. The need to navigate both roles, which are governed by distinct norms, impacts how individuals conduct themselves online, and presents methodological challenges for researchers. We conduct a systematic exploration on how an organization's official Twitter accounts and its members' personal accounts differ. Using a climate change Twitter data set as our case, we find substantial differences in activity and connectivity across the organizational levels we examined. The levels differed considerably in their overall retweet network structures, and accounts within each level were more likely to have similar connections than accounts at different levels. 
  We illustrate the implications of these differences 
  for applied research  by showing that the levels closer to the core of the organization display more sectoral homophily but less triadic closure, and how each level consists of very different group structures.
  Our results show that the common practice of solely analyzing accounts from a single organizational level, grouping together all levels, or excluding certain levels can lead to a skewed understanding of how organizations are represented on social media. 

\end{abstract}



\maketitle

\section{Introduction}

Social media platforms are an area where individuals' public and private lives can spill into each other \cite{zhao2009HowWhyPeople}. As much as individuals attempt to avoid possible mishaps with the disclaimer that ``views are my own and do not reflect those of my employer'', problematic posts can lead to dismissal. In fact, many organizations have social media policies that govern how their employees behave outside of work. As an extreme example, Joe Biden has two Twitter (or X)
accounts, one as the president of the U.S. and one as himself. 
No one would reasonably consider tweets coming from the latter to be any less representative of what the U.S. president is saying, but there are clearly still differences in what gets tweeted from which account -- and in fact the U.S. Supreme Court has recently opined that whether the president's tweets are official or unofficial “may depend on the content and context of each” \cite{durkinricher2024WhatKnowSupreme}.  
How large and widespread are the differences in behavior between accounts in different organizational roles? Beyond organizational regulations, what explains these variations? How is the interplay between these considerations for most individuals who do not have work and private accounts?

Our study aims to answer what an organization is on social media. Social media, like other technologies, shapes and is shaped by organizational practices \cite{orlikowski1992DualityTechnologyRethinking, orlikowski2007SociomaterialPracticesExploring}. Social media use by organization members can blur organization boundaries and challenge hierarchies and communication policies but also be affected by the formal organization boundaries, hierarchies, and policies. Using climate change policy communication on Twitter as our case study, we explore how the formal roles within an organization, or organization levels, are related to social media behavior.  

Understanding how organizations behave at different levels allows for different, perhaps more sophisticated, research questions and method choices, and may lead to new results in many disciplines. Some questions about organizations that have been studied using their social media behavior include, for example: 
1) How do organizations communicate and connect with each other to further their interests \cite{vu2020leads}? 
2) Do different organization identities lead them to strategically choose different activity levels and connections \cite{li2021organizational}? 
3) How does the authority of an organization develop and how does it influence policies \cite{goritz2022international}?
4) How do organizations manage their public relations with their stakeholders \cite{rybalko2010dialogic}?
5) What is the relationship between an organization's online presence and its economic resources \cite{whitesell2023local}?

Across different fields, research using social media to study organizational behavior tend to focus on official accounts (e.g., communication \cite{vu2020leads,li2021organizational}, policy studies \cite{hayes2018multiplex,goritz2022international}, management \cite{rybalko2010dialogic}, political science \cite{whitesell2023local,hemphill2013}, and computer-aided communication and cooperative work \cite{lovejoy2012,hemphill2013}). If organizations behave differently at various levels, previous findings may need re-examination. Using only official accounts can lead to sparse networks and an incomplete picture of the organization's online presence. It may overrepresent dominant mainstream voices, neglecting the potential of social media to amplify diverse and marginalized perspectives \cite{van2016weapon}. For instance, nonprofit organizations with limited budgets for official social media management might have passionate members actively using personal accounts to represent the organization. Studying only official accounts could lead to incomplete or erroneous conclusions about an organization's social media influence.
However, if there are no significant differences in the behavior of different levels of an organization, it is possible to decide which organization level to include based only on the amount of data required. 

We explore the implications of different organizational boundaries on social media to inform researchers' decisions when collecting and aggregating organization-related accounts. Our study aims to clarify the trade-off between data collection costs and capturing the full range of an organization's beliefs, behaviors, and norms. We hypothesize that main official accounts closely represent an organization's rules and positions, whereas the personal accounts of non-key individuals may differ significantly. Other official accounts and personal accounts of key people fall somewhere between the two extremes.
Finally, we refine our analysis by further considering how behavioral differences across organization levels may vary between types of organizations (e.g., NGOs, corporations, political parties). This will allow us to speak to, for example, how non-profit, non-governmental environmental organizations (ENGOs) might attract and expect a greater alignment in climate change views across all levels compared to for-profit firms, a pattern that might result from those joining the ranks of ENGOs being more likely to emphasize collective goals (e.g., curbing global emissions)
\cite{wallis2021activism}.

Social media platforms have enabled members at all organizational levels to participate and express themselves in ways not possible through traditional media or official organizational websites \cite{lovejoy2012}. However, while these platforms offer the potential for a level playing field, previous research and theoretical frameworks in computer-supported cooperative work \cite{orlikowski1992DualityTechnologyRethinking,oberg2008HierarchicalStructuresCommunication,schmidt1994OrganizationCooperativeWork,starbird2019DisinformationCollaborativeWork} suggest that existing organizational structures may still lead to differences in social media interactions along formal roles. This tension between the democratizing potential of social media and the persistence of organizational roles and norms forms a central focus of our investigation.

More broadly, our study has important implications for research on organizations across social sciences. As noted above, fields such as communication, policy studies, and management are interested in questions about organizations in social media. Which accounts are included as representatives of organizations may affect the answers to those questions. The answers in turn can have social implications if they affect, for example, social media strategies or even legal interpretations.

\subsection{Why organization level can matter in social media?}

Here we offer more detailed justifications for the exploratory hypothesis that the social media behavior of a member of an organization is affected by their role in the organization. We consider two reasons: incentives and social norms.

\subsubsection{Incentives}

People's behavior on social media can be affected by their own and their organization's economic and other incentives. This was implicitly acknowledged by a social media platform owner's promise to cover the legal costs of anyone who gets into trouble with their organization for their activity on his platform \cite{kim2023MuskVowsPay}. 

Social media accounts of an organization can differ in their perceived representativeness of an organization's views. An organization's political and economic interests may be advanced or harmed more by the official and the personal accounts of the executives than by the personal accounts of non-executive members. As a result, organizations may incentivize the behavior in social networks at different levels of organizations in different ways. Those with an executive role may also have more to lose if they are punished for their social media behavior than people in non-executive roles. At the level of official accounts, organizations can pay employees to handle the social media accounts. There may be explicit contracts with certain amount of activity required from such employees as well as strict guidelines for the content. This can lead to differences in behavior between the official and the personal accounts.

\subsubsection{Social norms}
In addition to legally binding rules and incentives in an organization's contracts, other social norms or expectations can also affect behavior in social media. There might be prescriptive norms that are not strongly enforced, but still affect behavior. Prescriptive norms refer to beliefs or statements about what behaviors and beliefs \emph{should} occur in the group. Descriptive norms may also have an effect as people tend to change their behavior to match the descriptive norms of their social group \cite{lapinski2005ExplicationSocialNorms,wallen2017SocialNormsMore,agerstrom2016UsingDescriptiveSocial}. Descriptive norms refer to behaviors and beliefs that \emph{are} commonly occuring in a social group.  Therefore, any existing differences between different groups within an organization can become stable and even more pronounced over time regardless of the original cause of the differences.

In the current study, the ties in the networks are climate change related retweets. The ties can be considered as endorsements of the views expressed in the original tweets \cite{metaxas2015retweets}. Today, many organizations tend to have official positions and goals related to climate change, which can lead to norms affecting whose climate change views are endorsed by retweets and how often. If everybody shares their organization's climate change views, it is possible that there are only few differences across the retweet networks when the organization levels are compared.

\subsection{Previous studies}

There have been no quantitative studies explicitly looking at the relationship between formal organizational roles and social media behavior across multiple organizations, but a number of related studies on differences between organizational roles suggest mixed possibilities when it comes to whether actors from different organization levels would behave differently on social media.

First, there is extensive evidence that actors within an organization differ considerably by their hierarchical position, not just in terms of formally mandated behaviors, but also in terms of their preferences, their outlook on work, and how they communicate. 
For example, Srivastava and Banaji \cite{srivastava2011CultureCognitionCollaborative} found that people differ in their expressed and their implicit support for norms about collaboration across organization levels, suggesting that hierarchical positions may shape both overt and underlying attitudes towards cooperative work.
Elsewhere, De Choudhury and Counts \cite{dechoudhury2013UnderstandingAffectWorkplace} found differences in affect expressions by job role and level when studying a company's internal microblogging tool. 
Further, as Öberg and Walgenbach \cite{oberg2008HierarchicalStructuresCommunication} found, these communication differences persist even in firms that explicitly endorsed norms of non-hierarchical communication structures, indicating that formal organizational structures may influence communication behaviors, even when attempts are made to flatten hierarchies.

Of particular relevance to our study, Stanojevic, Akkerman, and Manevska \cite{stanojevic2020GoodWorkersCrooked} found that experiences of voice suppression by supervisors at workplace were associated with more support for populist politics outside the workplace. The study shows that not only do actors within the same organization have different preferences, actors take advantage of alternative contexts to express these differences. 

Finally, we note that the picture is not uniform, as a number of exploratory studies of individual companies suggest that social media reduces the effects of hierarchical levels on communication \cite{riemer2015TopBottom,zhao2009HowWhyPeople}. Together, these mixed findings underscore the need for more comprehensive research, especially in the context of climate change communication where the topic itself is often polarizing.

There is a need for research that examines intra-organizational differences, particularly in the context of public-facing communication on topics like climate change. Our exploration of differences in social media behavior between organizational roles fills a gap in understanding organizational communication on social media and contributes to broader discussions about the interplay between formal structures and informal behaviors in social media. By focusing on climate change communication, we can examine how these organizational dynamics play out in a politically charged and sensitive context. The polarized nature of climate change discourse may amplify or alter the influence of organizational hierarchies on social media behavior, potentially revealing new insights into how organizations navigate complex public issues across different levels of their structure.

\subsection{Research questions}
Following our discussion of the importance of potential behavioral differences across different organization definitions to many disciplines, we pose three main questions that guided our research:

\begin{enumerate}
  \item \textbf{Differences between organization levels:} How are the formal organizational structures related to individual behavior in social media? Are there significant differences between organization's official accounts and member's personal account's or between the executive's accounts and the non-executive's accounts? Do noticeable variations exist in online activity levels and timings? Are there disparities in the underlying network structures such as connections within the organization and connections outside the organization across different roles?

  \item \textbf{Differences between organization types:} Do organizations differ in how the formal organization structures are related to behavior in social media? 

  \item \textbf{Implications to applied research:} What consequences arise from neglecting differences in online behavior between organizational roles when conducting applied research? How might overlooking these distinctions impact research findings?
\end{enumerate}

Using a climate change retweet network as a case study, we find substantial variations in activity and connectivity at different organizational levels moderated somewhat by organization type. The following sections describe the methods of data collection and network construction (Section~\ref{sec:methods}), the results that highlight activity patterns and network structures (Section~\ref{sec:results}), and the impications on applied research illustrated by Stochastic Block Models (SBMs) and Exponential Random Graph Models (ERGMs) (Section~\ref{sec:applications}). 

\section{Methods}\label{sec:methods}
In this section, we present details about our data collection protocol and network construction methods. The methods for the more specific questions are in Section~\ref{sec:results} for level difference exploration and in Section~\ref{sec:applications} for applications. As social norms often emerge from complex interactions among entities, whether individuals or collectives, in a specific context, rather than from top-down imposition, we use network-based methods that provide a general framework for analysing how structural properties of social media networks reflect social norms  \cite{wasserman1994social}.

\subsection{Climate Change Policy Network Data}
To explore our research questions, we use climate policy discussions on Finnish Twitter as our case study.
Specifically, we study the Twitter activity of known policy actors and organizations from the Finnish climate policy space \citep{chen2024climate}. The case of climate change policy making offers an ideal context for examining the differences in social media behavior across organizational levels for several reasons.

First, climate change represents an urgent, global social, economic, and political problem that involves a diverse range of stakeholders. These stakeholders include various types of organizations -- NGOs, companies, and government bodies -- each with different resources, strategies, and cultures for social media use. This diversity allows us to explore how organizational properties influence social media behavior across different levels, addressing the questions raised in our introduction about the nature of organizations on social media.
Second, climate change is a topic where personal convictions might not always align with organizational interests. This potential misalignment provides an opportunity to examine how individuals at different organizational levels navigate a possible tension between personal views and official organizational stances, a key aspect of our inquiry into the blurring of public and private lives on social media.
Third, the climate change debate is characterized by increasing politicization and polarization. In Western Europe, including Finland, an increasingly large share of citizens follow divergent cues concerning climate policy from different political parties. A similar trend of issue politicization and mass polarization has been developing in the US since at least the early 2000s \cite{fisher2022climate,smith2023partisanship}. This polarized environment allows us to explore how organizational hierarchies and individual agency interact in shaping social media discourse on a contentious issue.

In this increasingly politicized context, formal organizations need to carefully consider the interests of their stakeholders in the long term. They must pay particular attention to the ways, risks, and benefits of communicating about climate change on public forums such as Twitter \cite{falkenberg2022growing,frandsen2011rhetoric,figenschou2020interest}. This strategic consideration at the organizational level may contrast with the more immediate and personal communications of individuals within these organizations, potentially highlighting differences across organizational levels.

By focusing on climate change policy network, we can gain insights into how organizational structures influence social media behavior in a context where the stakes are high and the boundaries between personal and professional, public and private, are often blurred. This approach allows us to address the central questions of our study about what constitutes an organization on social media and how behavior differs across organizational levels.

\subsection{Data Collection Protocol}
The network actors were identified based on the knowledge of the actors that participate in climate policy making in Finland. A list of actors was made in consultations with climate policy experts, which resulted in a roster of 241 organizations that represent different sectors of society: governmental, scientific, business, civil society, media, interest groups, and political parties. After the bounds of the network were identified following \cite{yla2018climate}, the connections between its members were mapped with surveys and Twitter data collection following the protocol described in \cite{chen2024climate}. We collected our Twitter data using the Twitter application programming interface (API). Data from 2017--2021 was collected in December 2021, and data from 2022 was collected in February 2023. We focus on the years from 2017 through 2022 as Finnish Twitter was generally very quiet prior to 2017; the salience of climate change as an issue increased dramatically in 2018 following the publication of the so-called "1.5°C report" by the Intergovernmental Panel on Climate Change \cite{gronow2024partisan}. We included all tweets and retweets of the accounts in the raw data but restricted the data in the network to retweets that included substrings related to climate change.

The Twitter accounts were linked to organizations and they were assigned to one of four levels within each organization:
\begin{enumerate}
    \item organization’s main account (\textit{organization main}),
    \item organization's sub-divisions' accounts (\textit{organization side}),
    \item personal accounts of the organization’s top executive leaders (\textit{individual main}), or
    \item personal accounts of other members of the organization (\textit{individual side}).
\end{enumerate}

\subsection{Network Construction}\label{sec:network_construction}

We constructed two types of graphs: 
(1) an account-level graph, where each user account constitutes a vertex and edges represent retweets between them, and (2) collapsed graph, where original accounts are aggregated into organizations as vertices with backboned edges.
In collapsed graphs each organization is represented by a single vertex. The various accounts belonging to the organization are collected to that vertex.
The edges in this graph are between organizations, and two organisations are connected if there is a retweet between an account in one organization and an account within the other organization and if it is not removed by backboning.
The set of retweet edges in the collapsed graphs is trimmed and made binary by backboning which is explained in more detail below in Section \ref{sec:backboning}. 

The account-level graphs were used in estimating the connectivity within (Section~\ref{sec:density}) and outside (Section~\ref{sec:overlap}) organizations. The collapsed graphs were used in applications (Section~\ref{sec:applications}). When comparing the levels of organizations, the graph was transformed into a subgraph of a particular organizational level. Only accounts of that level were included as vertices, and only retweets between the accounts of that level were considered when constructing the edges. The structural comparisons and the applications also used null models or bootstrapping to estimate randomness in the estimates and backboning to reduce noise. These methods are descibed below in Section~\ref{sec:methods-bootstrapping}. 

We studied activity differences among accounts by looking at all of their tweets (i.e. original tweets, replies, retweets, and quote retweets) during the time period. When constructing the networks, we only used climate change related retweets. In account-level graphs, we assigned weights to the edges as the number of retweets between the two accounts. We focus on retweets because they are a commonly studied behavior for Twitter networks, and as generally unconditionally positive interactions \cite{metaxas2015retweets}, they are an appropriate measure of shared values or for the application to policy studies, which tend to focus on cooperative relationships, than other types of Twitter interaction (e.g. quote retweets) which tend to be used for more varied signals \cite{garimella2016quote}. The resulting networks can therefore be considered as policy endorsement networks.

\subsubsection{Backboning}
\label{sec:backboning}
Weighted networks, such as ours in which each edge has a weight attribute to indicate the number of retweets between a pair of Twitter accounts, are multidimensional data structures that are good at representing complexity, but often contain noise (e.g., atypical behavior during an exceptional event) that obfuscates the core structure of the network. To tease out the core structure of the weighted network while facilitating meaningful insights into the underlying data, researchers typically perform dimension reduction by extracting the "backbone" of the network, in which each edge indicates the presence of a  statistically significant connection \cite{neal2022backbone}.

The choice of the specific technique depends on the properties of the original weighted network (e.g., clustering coefficient and edge weight distribution), but for our retweet networks, we extracted the so-called noise-corrected backbone that identifies the strength of connections among vertices by accounting for both the edge weights and the patterns of connections \cite{coscia2017noisy}.  The algorithm goes through every edge in the network discerning which edges are significant under a null model, emphasizing the most relevant connections while mitigating the effect of noisy or less substantial interactions. The method ultimately ranks the edges by their significance and allows for removing edges exceeding a certain significance level, $p \geq 0.1$ in our case. In addition to placing emphasis on the statistically significant connections, the resulting binary, unweighted networks enable the use of standard network analysis methods.

\subsubsection{Bootstrapping}
\label{sec:methods-bootstrapping}
In addition to backboning, we estimated the effects of randomness on the network structures by bootstrapping. Specifically, we used bootstrapping to control for the possible effects of the activity level or the size of the network when comparing different organization levels with each other. 

The method consisted of, first, calculating the probability of each edge from the observed retweet graphs by dividing the edge weight by the total graph weight, and, second, taking random samples from the edges based on the edge probability distribution and on a given graph size. The edge sample size could be the same as the size of the graph used in calculating the observed edge probability distribution, or it could be the size of another graph that we were interested in. We then used the drawn samples of the same size to compare different organization level networks with each other.  

\section{Results} \label{sec:results}
Having described our overall empirical approach and data collection protocol, we now turn to examining the differences in social media behavior across organization levels. We proceed by first presenting descriptive statistics of our network data. We then analyze activity patterns and examine the network structure across organizational levels without considering organization types. Finally, we delve deeper by exploring how these patterns and structures vary not only across organizational levels but also across different types of organizations.

\subsection{Descriptive Statistics}

Table~\ref{tab:descriptives} presents the descriptive statistics of the two types of networks that we built and used, account-level graphs and collapsed graphs. There were 8 384 accounts affiliated with the 241 roster organizations in total. The retweets between accounts are during the time period from January 2017 to December 2022. For the account-level graphs, we built graphs whose vertices were accounts with a certain organization role. For the application analysis, we used collapsed graphs where the vertices were organizations to which affiliated accounts were added for each level after trimming the edges between the  account-level vertices to binary backbones. It was possible that some organization vertices in a collapsed graph had no accounts or edges associated with them. The graph with the accounts from all levels collapsed to their organizations is visualized in Figure~\ref{fig:collapsed_network}.

When comparing the graphs by levels, the fewest vertices were in the graph consisting of all organization main accounts. The most vertices were in the graph with all individual side accounts, which also had the most retweets or total edge weight. The fewest edges were in the graph where organization side accounts were collapsed into single organization vertices. As we will see later, this last graph (collapsed and backboned organization side) produced the widest uncertainty estimates as well as the least amount of community structure.

\begin{minipage}{\textwidth}
\centering
    \begin{minipage}{.65\textwidth}
    \centering
        \begin{tabular}{lrrrrr}
            \toprule
            & \multicolumn{3}{c}{Account-level graph} & \multicolumn{2}{c}{Collapsed graph} \\
             \textbf{Level} & \textbf{Vertices} & \textbf{Edges} & \textbf{Weights} & \textbf{Vertices} & \textbf{Edges}\\
             \midrule
             all levels, directed & 8 384  &  60 283   & 218 751  & 241   & 6220      \\
             organization main      &  209   &  1 294    & 9 718    & 241  &  625      \\
             organization side      &  916   &  474      & 1 682    & 241  &  70      \\
             individual main      &  802   &  2 624    & 10 641   & 241  &  542     \\             
             individual side      &  6 457 &  16 735   & 46 153   & 241  &  732    \\
            \bottomrule
        \end{tabular}
        \captionof{table}{Descriptive Statistics of the Observed Graphs}\label{tab:descriptives}
    \end{minipage}%
    \begin{minipage}{.3\textwidth}
        \centering
        \includegraphics[width = 1.0\textwidth]{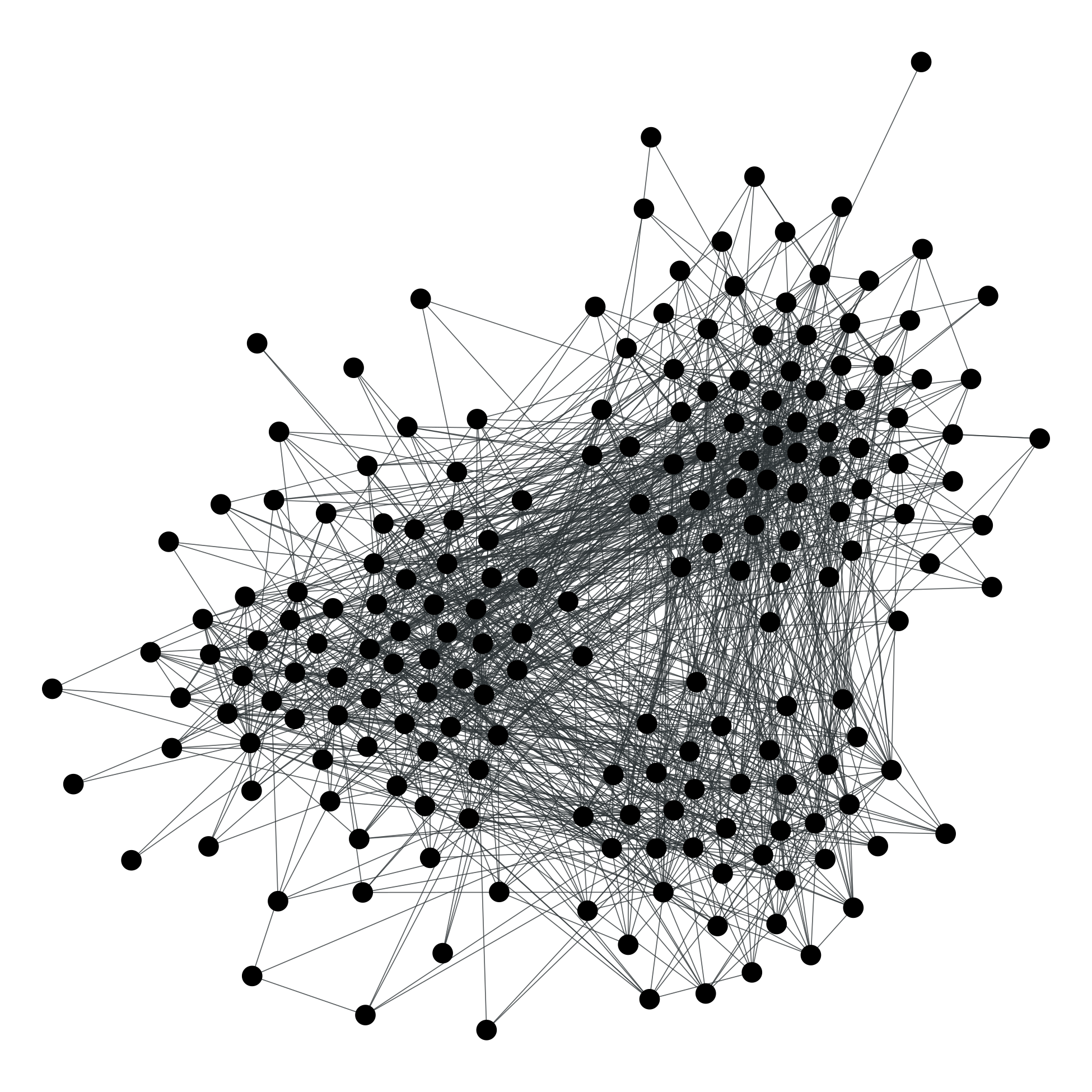}
        \captionof{figure}{Collapsed Graph with All Levels}\label{fig:collapsed_network}
    \end{minipage}
\end{minipage}

\subsection{Activity} 

\subsubsection{Weekly and Daily Patterns}

\begin{figure}[ht]
    \centering
    \includegraphics[width=1\textwidth,trim={40pt 0 60pt 30pt},clip]{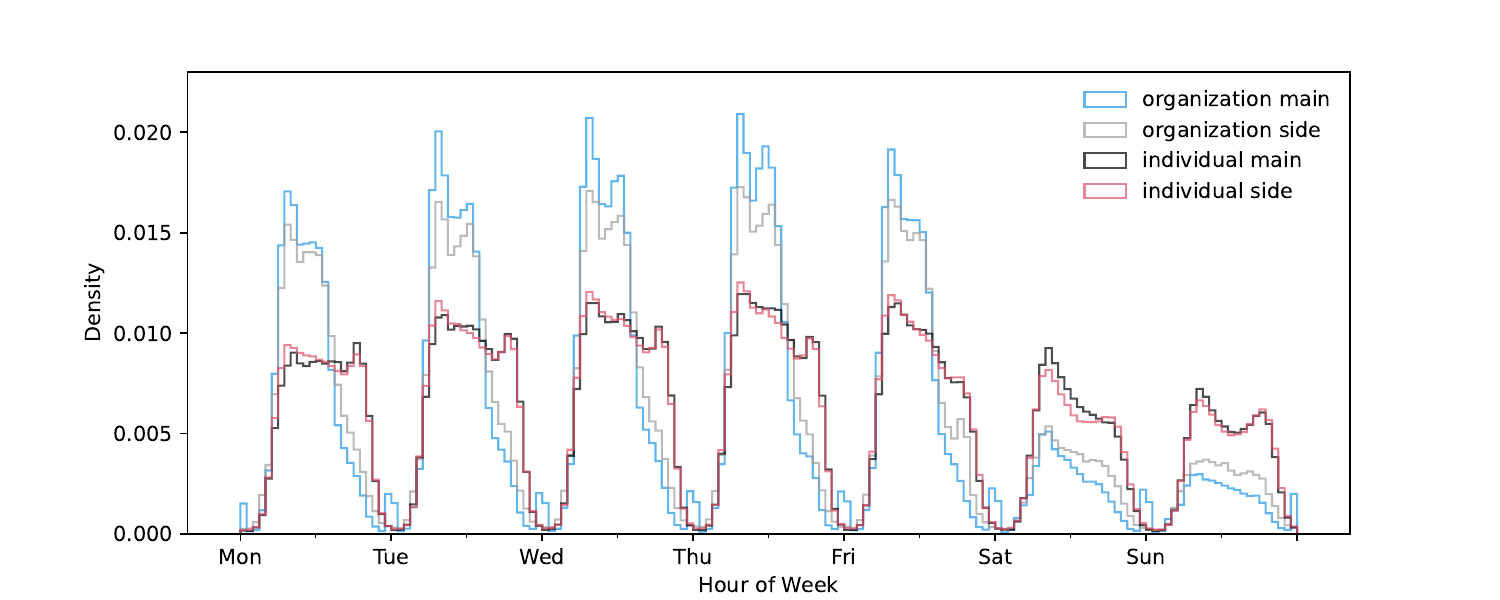}%
    \caption{Activity Level by Hour of Week} 
    \label{fig:activity}
\end{figure}

The normalized account tweeting activity for the hour of the day of the week is shown in Figure~\ref{fig:activity} for the four organization levels. The official organization accounts tended to tweet more during office hours and work days than at other times. The decrease of the activity outside office hours and during weekends was smaller for the individual accounts. The pattern is consistent with official organization accounts being run by paid personnel who use them mostly during office hours for work, and the personal individual accounts used in a more similar manner during work days and weekends because they are also used for non-work related purposes.

There is a prominent regular peak for the organization accounts in the mornings of work days which may reflect scheduled and programmed tweets by the organization accounts. The question of whether the activity is done manually by humans or whether it is scheduled can be studied by examining the burstiness of the accounts.

\subsubsection{Burstiness predictions}

Individuals do not in general engage in activities, such as posting in social media, in regular or even uniformly random time intervals, but the activity is typically more bursty \cite{karsai2018bursty}. However, organizations that have multiple users for the same account could exhibit more random temporal patterns, and social media strategies or professional social media use could even exhibit more regular patterns. 

We use the bustiness coefficient to measure how regular or bursty the activity of individual accounts \cite{goh2008burstiness}. The coefficient is defined as
\begin{equation}
    B = \frac{\sigma_\tau- \mu_\tau}{\sigma_\tau+ \mu_\tau}\,,
\end{equation}
where $\mu_\tau$ is the average time between posts of the user and $\sigma_\tau$ is the standard deviation. The coefficient takes value $-1$ for completely regular sequence, $0$ for completely random sequence (i.e., one produced by a Poisson process), and $1$ for maximally bursty sequence.

\begin{figure}[h]
    \begin{minipage}[b]{0.45\textwidth}
        \begin{tabular}[b]{ccrc}
            \toprule
            \textbf{Group 1} & \textbf{Group 2} & $\textbf{\textit{M}}_1 - \textbf{\textit{M}}_2$ &\textbf{CI [2.5\%,  97.5\%]} \\
            \midrule
            org main & ind main & $-0.038$ & $-0.094, 0.019$\\
            org main & org side & $-0.001$ & $-0.056, 0.054$ \\
            org main & ind side & $-0.012$ & $-0.064, 0.040$ \\
            \textit{ind main} & \textit{org side} & $0.037$ & $\hphantom{-}0.006, 0.067$ \\
            \textit{ind main} & \textit{ind side} & $0.026$ & $\hphantom{-}0.000, 0.051$  \\
            org side & ind side & $-0.011$ & $-0.032, 0.010$ \\
            \bottomrule
        \end{tabular}
        \captionof{table}{Burstiness Comparison: Tukey's HSD}
        \label{tab:burstiness_comparison}
    \end{minipage}\hfill
    \begin{minipage}[b]{0.45\textwidth}
        \includegraphics[width=\textwidth]{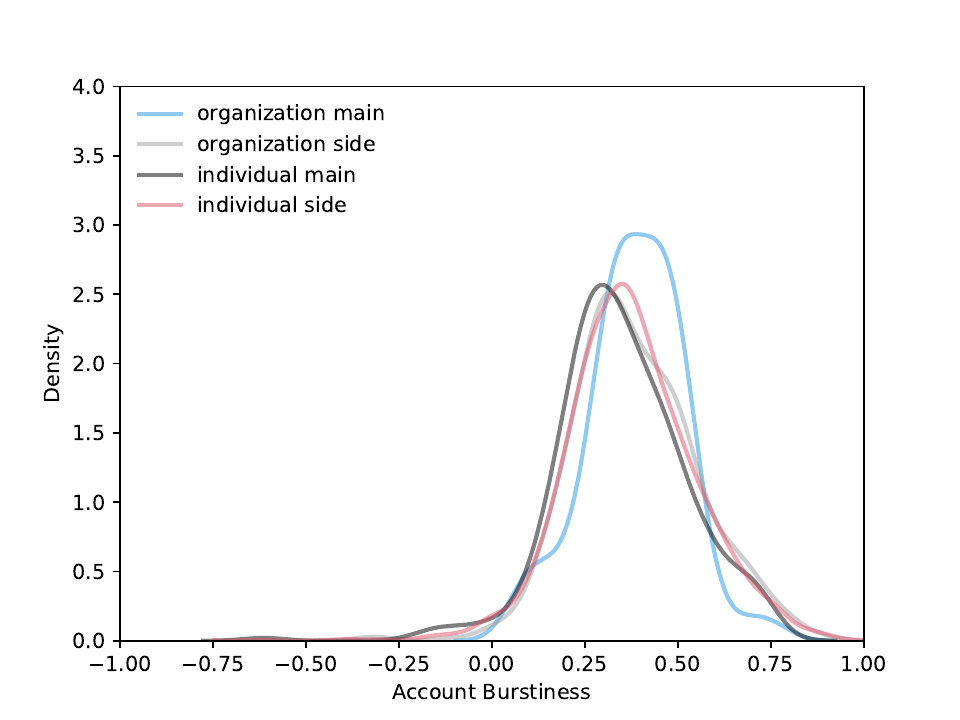}
        \captionsetup{singlelinecheck=false, justification=centering}
        \captionof{figure}{Burstiness Density}
        \label{fig:burstiness_comparison}
    \end{minipage}
\end{figure}

The distributions of burstiness of different organization levels are shown in Figure~\ref{fig:burstiness_comparison}. The differences across levels are not large. The burstiness estimates tend to be around values that are expected from human users (0.2 -- 0.6) \cite{gandica2016OriginBurstinessHuman}. We tested whether the burstiness means differed across levels using Tukey's Honestly Significant Differences (HSDs) test (Table~\ref{tab:burstiness_comparison}). We found statistically significant difference at \textit{p} < 0.05 level for individual main account in comparison to organization side and individual side accounts. However, the sizes of the differences were insignificant in practice. Although the organization accounts exhibit different activity patterns than the personal accounts as revealed by activity by hour of week, in aggregate, the accounts at all levels appear to be used mostly manually by humans.

\subsection{Network structure}
We study the relationship between organisation level and network structure by two basic questions: How likely are the different organisation levels to connect within themselves, and how likely are the accounts within an organisation level to have similar connections overall? We operationalize both questions by studying mixing matrices (Section~\ref{sec:mixing-matrices}), and the first question by computing the density (Section~\ref{sec:density}), and the second one by overlap (Section~\ref{sec:overlap}).

\subsubsection{Mixing matrices by organization level}
\label{sec:mixing-matrices}

We employ mixing matrices to analyze the probabilities of endorsing others' views within and between organizational levels, irrespective of whether other accounts are part of the same organization. These matrices consist of accounts that are not collapsed to their respective organizations, and endorsement is measured with directed connections with any account other than itself. 

\begin{figure}[h]
    \centering
    \includegraphics[width = 0.5\textwidth]{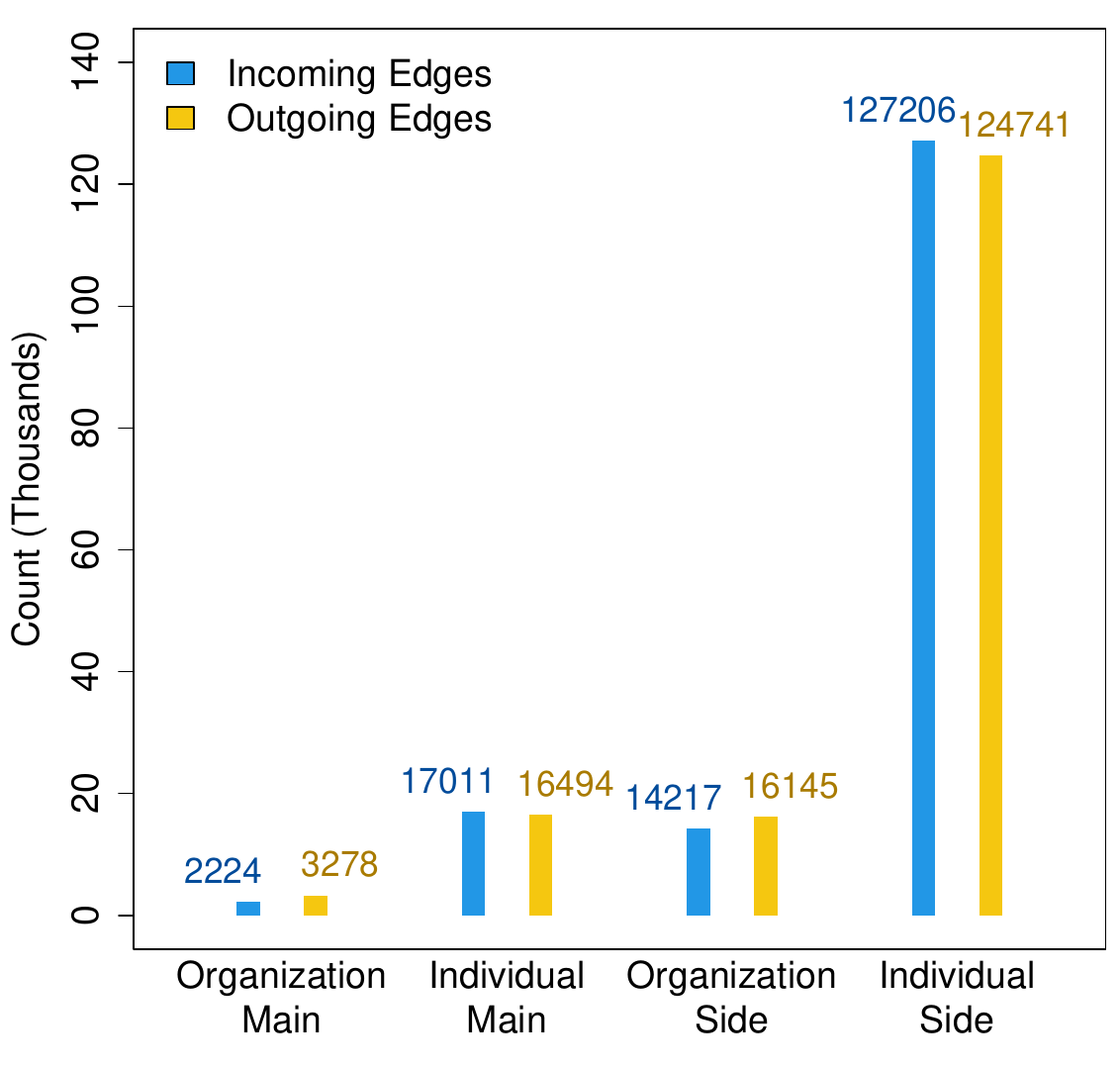}%
    \includegraphics[width = 0.5\textwidth]{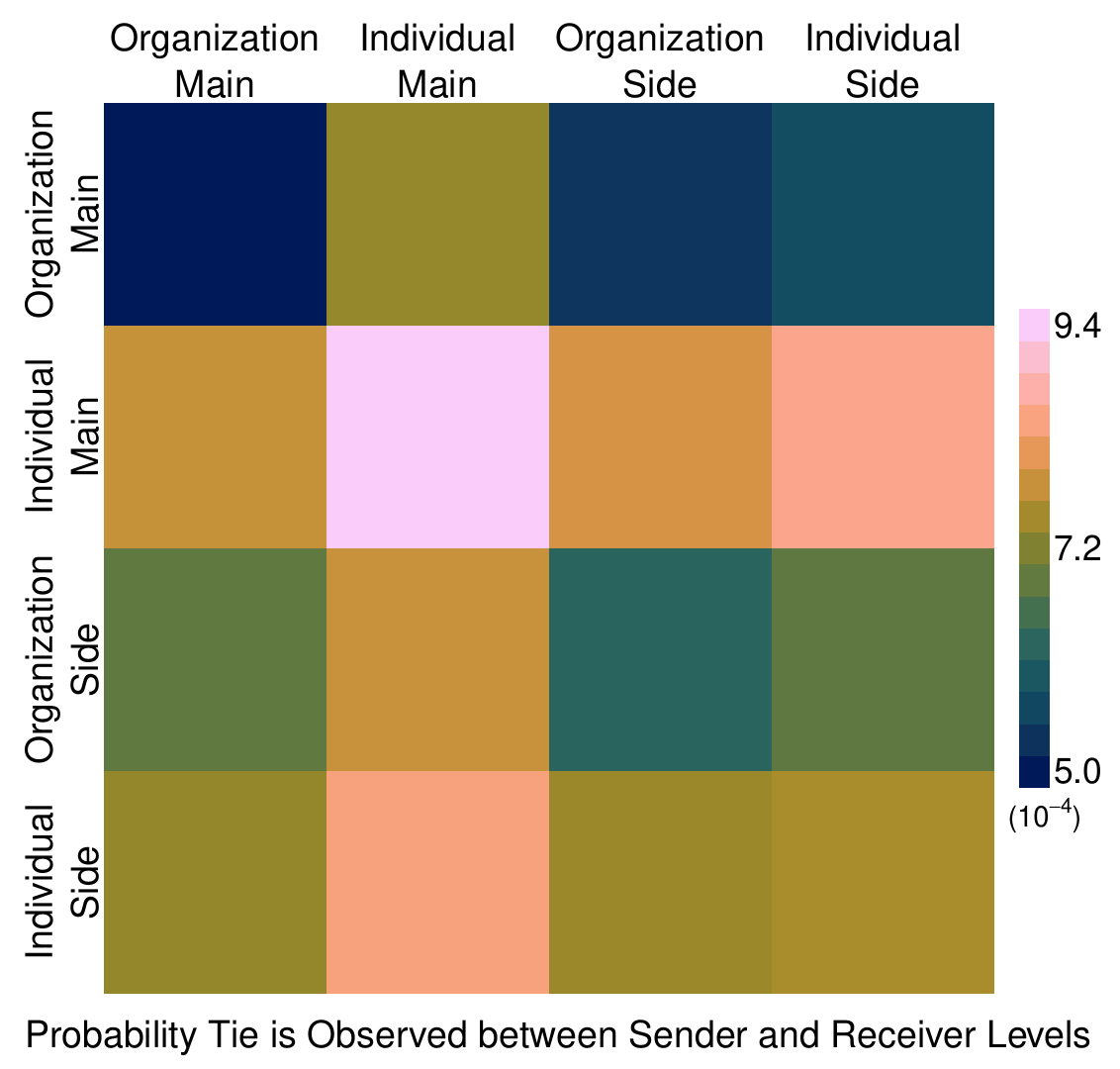}%
    \caption{Mixing Matrices for Organization Levels. Cells represent probability of observing an edge for the given sender type (row) and receiver type (column).}
    \label{fig:mixing-matrix}
\end{figure}

Figure~\ref{fig:mixing-matrix} visualizes the probabilities of endorsements between the organizational levels. Substantial differences exist between organization levels in terms of whose views are endorsed. Considering the baseline probability of observing an endorsement between a certain level-pair, shown by cells which are counts of endorsement edges normalized by the number of directed dyads of that level-pair, we see that two accounts from the individual main level has the highest probability of endorsing each other. On the other hand, the least likely observed endorsement is between two accounts from the organization main level. While the absolute probability of observing these endorsements is very low, the individual main level-pair is 1.5 times as likely as the organization main level-pair to have such a connection.

The comparisons show strong asymmetries in endorsements. Individual side accounts appear to play a central role in receiving and broadcasting endorsements across different organizational levels. Individual main accounts, on the other hand, show more cohesion and mutual endorsement than others. Interpretively, this finding aligns with the notion of interlocking directorates, which describes a network phenomenon where directors of organizations serve simultaneously on the boards of other organizations, facilitating coordination and information flow among organizations involved \cite{bonacich1986interlocking}. Alternatively, executives could simply be closer to each other (i.e., fewer steps away) in offline social networks, the effect of which could be amplified by the relatively small size of the executive class in Finland. Yet, social norms concerning either social hierarchy at the level of the entire society (a descriptive norm) or the avoidance of signalling favoritism (a prescriptive norm) could also prevent executives from endorsing members from a lower organization level.


\subsubsection{Densities}
\label{sec:density}
The number of links within accounts of an organization measures the extent to which they endorse each other and collaborate in distributing each others messages. Parts of retweet networks with high densities have been associated with shared opinions or attitudes and density based methods are commonly used to identify polarised groups and opinion bubbles \cite{garimella2018quantifying,salloum2022separating,falkenberg2022growing}. Here, densities can be interpreted as the extent of endorsement of an organization's views about climate change.

The density measure we used was the share of of observed edges out of all possible edges between an organization's accounts. 
We calculated the density for each organization using the original accounts and retweets and then calculated the mean of those organization densities over all organizations. To compare the levels, the organizations were restricted to consist of accounts from only each specific level in that organization. To estimate the effects network size on the results, we also calculated the overlaps for 300 boostrapped networks for each level where the network size was restricted to the size of the individual main graph (boostrapping described in Section~\ref{sec:methods-bootstrapping}). Organization main level was not included to the comparisons because organizations usually had only one such account.

\begin{figure}[ht]
    \centering
    \includegraphics[width=0.5\textwidth]{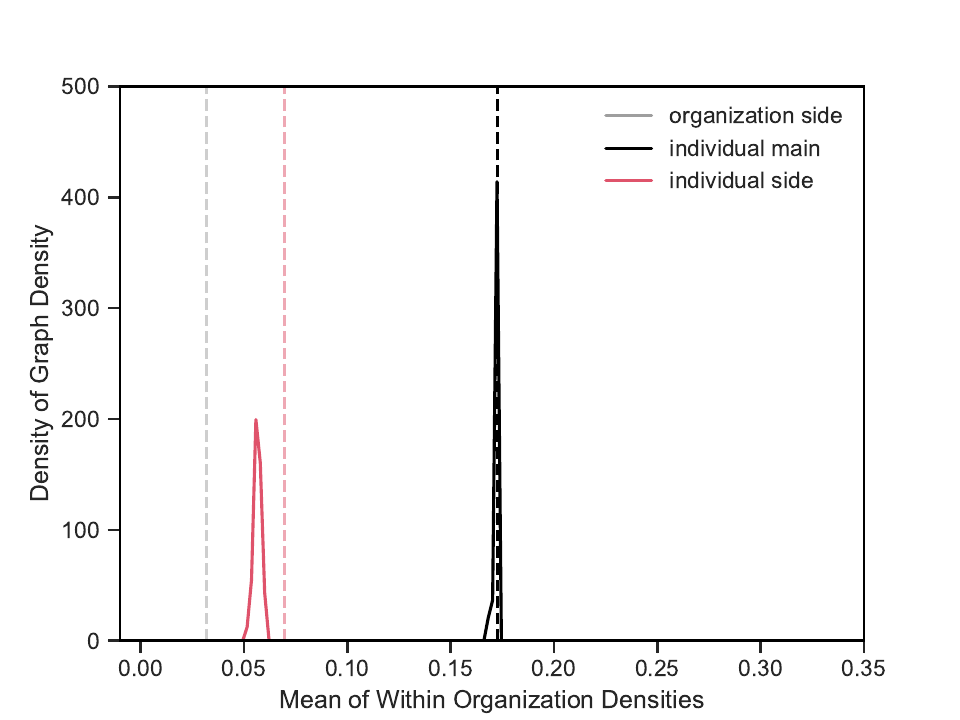}%
    \includegraphics[width=0.5\textwidth]{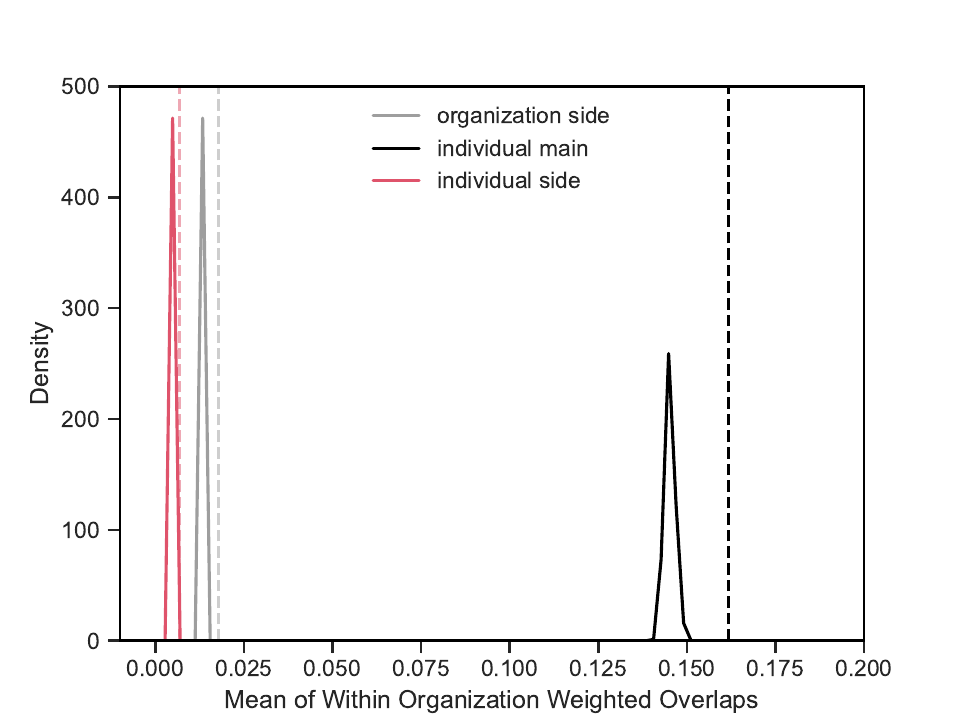}
    \caption{Within Organization Densities (left) and Weighted Overlaps (right)}
    \label{fig:densities_and_overlaps}
\end{figure}

As shown in Figure~\ref{fig:densities_and_overlaps}, both the observed (broken lines) and the bootstrapped mean densities (solid line distributions) differed by level. Individual main accounts had the highest density, a bit less than 0.2, and organization side accounts had the lowest density, close to zero. Individual side density was closer to organization side density than individual main density. In other words, the personal accounts of the executives tend to endorse more of the climate change related views existing in the organization than accounts with other roles. This strengthens the mixing matrix result about interlocking directorates (Section~\ref{sec:mixing-matrices}).

\subsubsection{Overlaps}
\label{sec:overlap}
We study to which extent accounts share the climate change views of the same people by measuring the similarity of their network neihgborhood. This is a slightly different question than was answered by density which measures the connectivity only within an organization. If two accounts would share the views of the same people, they have very similar neighborhoods in the network, i.e., they would be retweeting (and not retweeting) content from same users and their content would be retweeted (and not retweeted) by the same set of users. We quantify to which extent this is true by measuring the similarity of the neighbourhoods of pairs of users in the network. 

In order to account for the overall activity of the accounts we normalise the shared neighborhood size with the sizes of the neighborhoods of the two vertices. Computing similarity of neighborhoods of vertices has been explored in the context of quantifying the similarity of social networks of connected vertices, and within this literature several possible ways of doing the normalisation have been identified \cite{granovetter1973strength,onnela2007analysis,urena2020estimating}. We adopt the link overlap method \cite{onnela2007analysis,mattie2017generalizations}, but use it for all pairs of vertices within each organisation. The weighted neighborhood overlap \cite{mattie2017generalizations} measures how much of the retweet activity is common with two accounts giving more weight to overlap between neighbors which are at the end of high-weighted links:

\begin{equation}
    O_{ij}^W = \frac{\sum_{k \in n_{ij}} (w_{ik} + w_{jk})}{s_i + s_j + 2w_{ij}},
\end{equation}
where $n_{ij}$ is the set of common neighbors of vertices $i$ and $j$, 
$w_{ij}$ denotes the weight associated with the edge between vertices $i$ and $j$, and $s_i$ ($s_j$) denotes the sum of all edge weights associated with of vertex $i$ ($j$). One can also define an unweighted overlap measure, see Section~\ref{sec:woverlap}. 

From Figure~\ref{fig:densities_and_overlaps} we can see that the mean unweighted neighborhood overlaps are fairly small for both personal account levels and for the organization side accounts. Similarly to the densities, the personal accounts of those with an executive role have more common connections than accounts at other organizational levels. The personal accounts of the executives therefore not only share climate change views with more people within organizations than other accounts, as measured by densities, they also share more of the same people's views, as measured by neighborhood overlaps. 

There are clear differences across organization levels in their activity and network structure. These differences may have significant implications for the analyses of organizations in social media. 

\subsection{Organizational Levels Across Organization Types}
\label{sec:organization-type-results}
\begin{figure}[ht]
    \centering
    \begin{tabular}{@{}c@{\hspace{-0.2cm}}cccc@{}}
        \includegraphics[height=8cm, width=1.5cm]{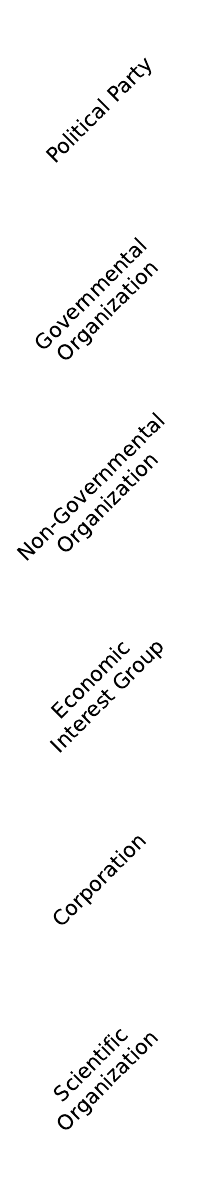} &
        \includegraphics[height=8cm]{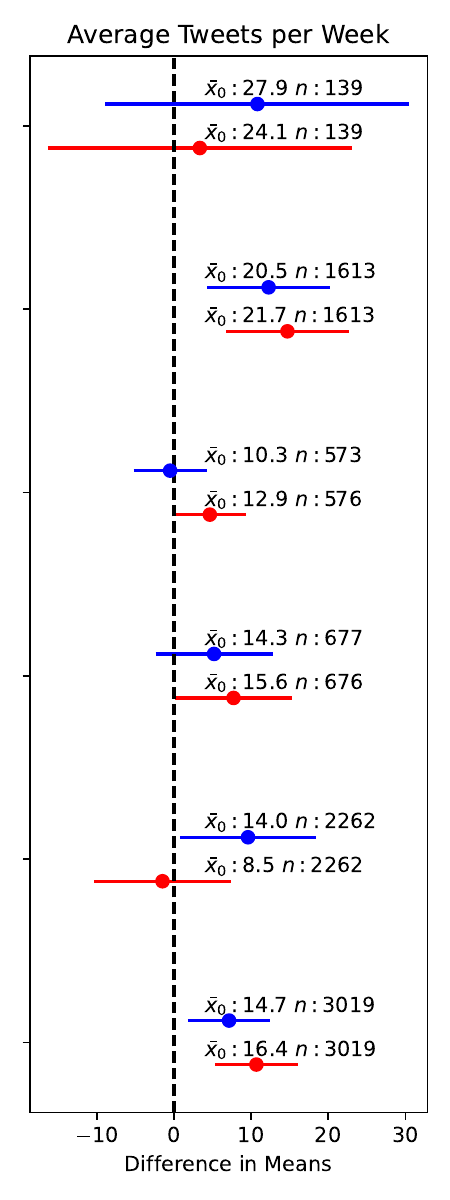} &
        \includegraphics[height=8cm]{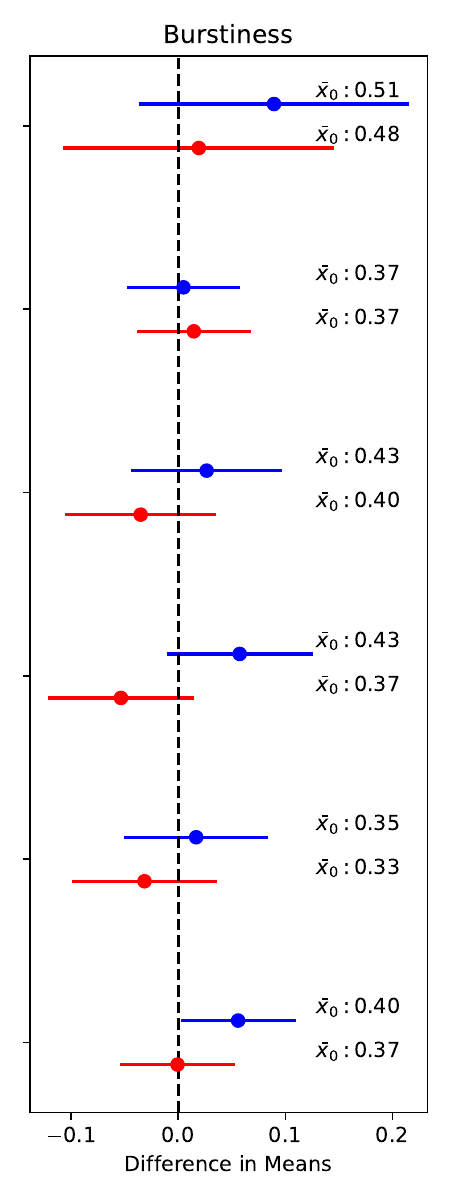} &
        \includegraphics[height=8cm]{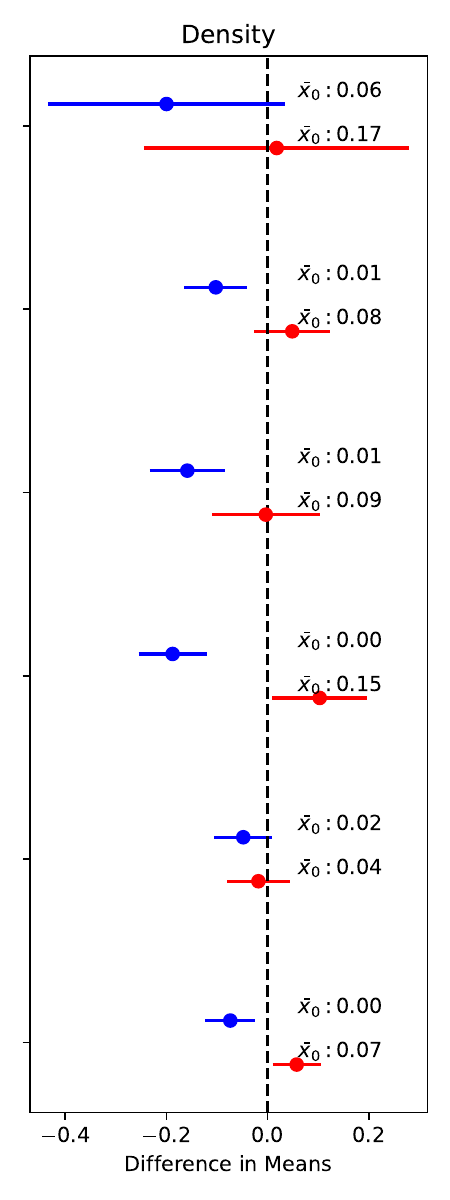}
        &
        \includegraphics[height=8cm]{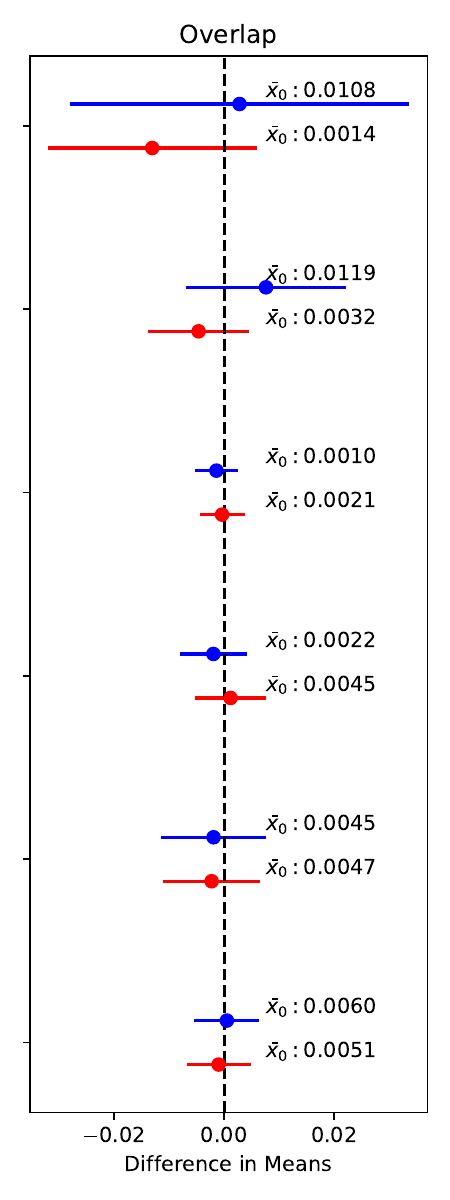}
    \end{tabular}
    \includegraphics[width=\textwidth]{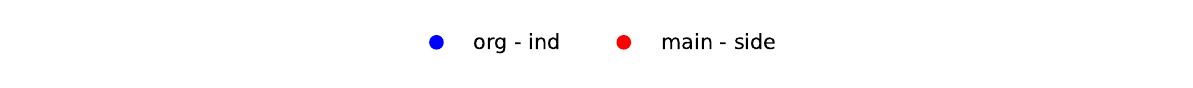}
    \caption{Differences in activity and connectivity by organization level and type. The circles represent the differences of the means. The lines from the circles represent the 95\% confidence intervals. $\Bar{x}_0$ is the mean of the first level aggregation in the difference and $n$ is the number of accounts in both of the compared level aggregations combined. The overlap measure is the weighted version.}
    \label{fig:levels-by-type}
\end{figure}

As discussed in the introduction, organization level differences might not be the same for all types of organizations. For example, Jacobs and Watts \cite{jacobs2021LargeScaleComparativeStudy} found large differences among 65 firms in communication activity and informal social network structure. Those difference can extend to how people behave across levels. We examined differences between six types of organizations: political parites, governmental organizations, non-govermental organizations, economic interest groups, corporations, and scientific organizations (Figure~\ref{fig:levels-by-type}). We looked at the differences between means of organization and individual accounts as well as main and side accounts.  

In general, the differences across organization types are not large. Non-governmental organizations have the smallest difference in the activity between the official and personal accounts. Corporations, and governmental and scientific organizations have more active official accounts than personal accounts on average. 

The burstiness values are within values typical for human activities \cite{goh2008burstiness} across different organization types and levels. 
Surprisingly, the organization accounts seem a bit more bursty than individual accounts. Side accounts are estimated to be a bit less bursty than main accounts for NGOs, economic interest groups, and corporations.
The burstiness differences between levels are smallest among governmental organizations, and largest in the economic interest groups. 

The differences in densities follow a  similar pattern across organization types. For all organization types, the personal accounts are clearly more densily connected within organizations than official accounts, which have densities close to zero.
This difference is
less pronounced 
 for corporations and scientific organizations, which still maintain close to zero official account density. The main accounts are more densely connected than side accounts in scientific organizations, economic interest groups, and governmental organizations. There do not appear to be large differences between levels in corporations using this aggregation. 

The overlap values are overall very small across different organisation types and levels, which means that the organisation type and level is not a good predictor for the interaction patterns of an account. 
This is in contrast to the larger overlaps in the individual main accounts noted previously which are specific to that level and not common to all main or all individual accounts.

\section{Applications} \label{sec:applications}
Now that we have shown that there are differences in how different organization levels behave on social media, we turn to see how these differences, and not accounting for them, substantively impacts applied research.
We illustrate the implications of these differences with two common types of applied analysis: stochastic block modelling, which is used to study social or group sorting behavior, and exponential random graph modelling, which is used to study tie formation between social actors. We demonstrate how, and when, using different organization levels for these two common tasks can lead to different conclusions on the group structure present in the network and on the presence and strength of different social network evolution mechanisms in the networks. 


\subsection{Stochastic Block Modeling}
The existence of groups and how they are formed is central to many social scientific questions. In political science, much attention is paid to how groups of political actors emerge to compete for institutional, economic, coercive, discursive, or other forms of power \cite{malkamaki2023complex}. 
For example, persistent north-south disparities among NGOs within a transnational advocacy network were uncovered by looking at the groups emerging from the relatively local interactions among them \cite{cheng2021advocacy}. 
Much of the contemporary polarization research also relies on identification of structural groupings within social or political systems \citep{salloum2022separating}.
Within retweet (i.e., endorsement) networks on climate change topics, previous work has found highly connected groups, which have been interpreted as evidence of polarisation at the level of individual accounts \cite{chen2021polarization,xia2021spread,falkenberg2022growing}. 
However, as we discussed, most of these analyses are usually done without critical consideration of how the organizational levels studied impacts their conclusions.

\begin{figure}[!t]
    \centering
    \includegraphics[width = \textwidth]{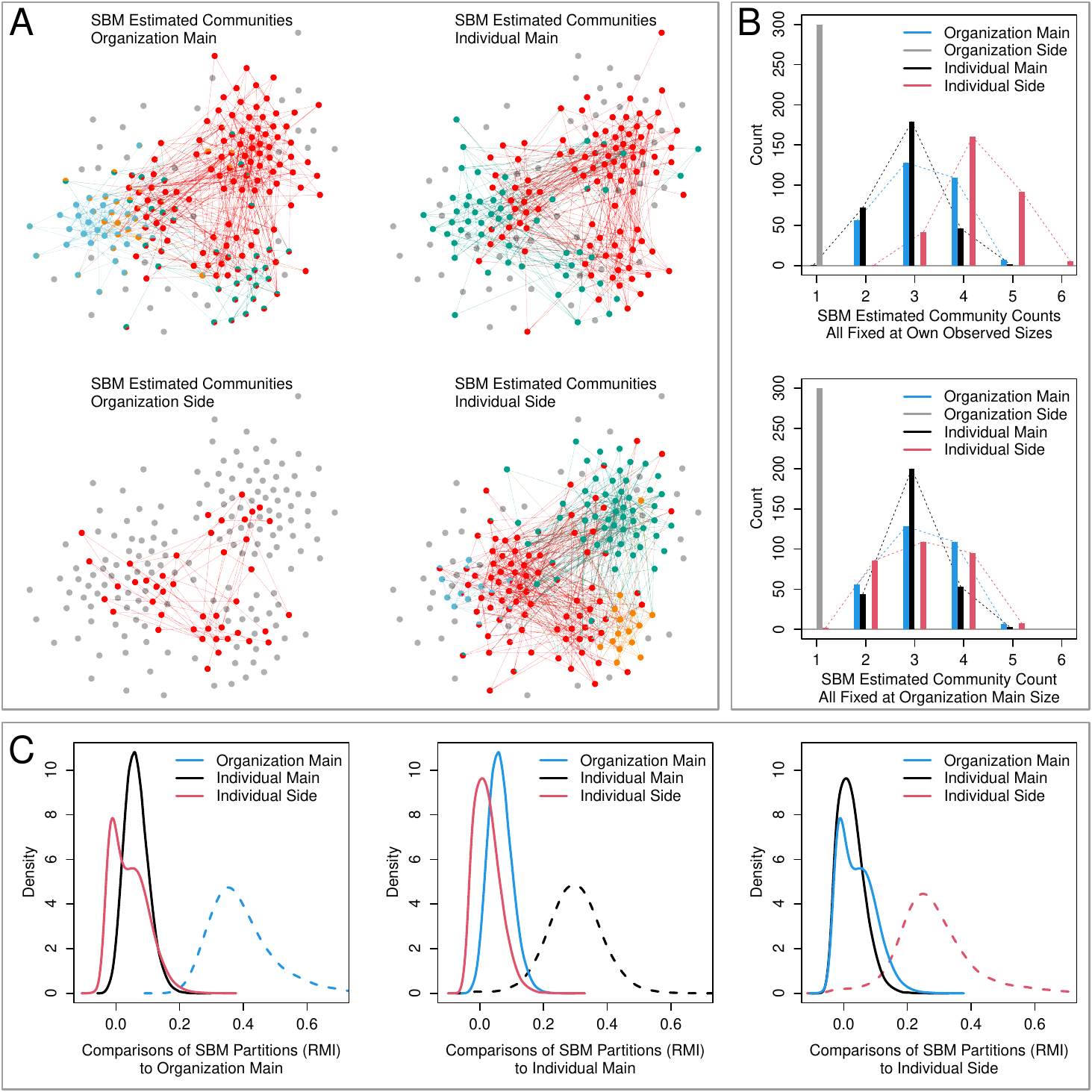}
    \caption{Panel A: Community Assignments for the Observed Networks. Vertices are collapsed organizations. Gray vertices indicate the organization is not observed in the given level, while all other colors are community assignments. Panel B: Number of estimated communities, bootstrapped network sizes fixed to the levels' observed size (top) and to the size of organization main (bottom). Panel~C:~Similarity between SBM communities in terms of RMI. Dashed lines shows similarity between bootstrapped networks of the same level and solid lines show similarity across bootstrapped networks from different levels.}
    \label{fig:snorms_assignments}
\end{figure}

To examine how choosing a particular organization level alters our understanding of such emergent behavior on social media, we applied the same group structure analysis to networks constructed from different organization levels to see how the results differ. The analysis used in these studies typically attempt to characterize the group structure in social networks by algorithmically dividing the vertices of a graph into blocks based on how they are connected by edges they have formed with one another \cite{lorrain1971structural,holland1983stochastic,porter_communities_2009}.
This produces a block structure -- the division of vertices into subgroups --
such that the blocks capture substantively meaningful group memberships.
For this application, we specifically fitted nested and degree-corrected stochastic block models (SBMs) -- a nested SBM is a generative model that hosts a block structure while accounting for the possibility of a hierarchical structure \cite{karrer_stochastic_2011,peixoto_hierarchical_2014} -- and compared the resulting block structures across organization levels. In our analysis, model parameters were inferred from the data using a merge-split Markov Chain Monte Carlo (MCMC) algorithm \cite{peixoto_merge_2020}.

We begin showing our results as the inferred block structure in Figure~\ref{fig:snorms_assignments}A. With each of the collapsed graphs plotted using the same layout, we are able to see a striking difference across the organization level in terms of the community assignments. First, we note that few organizations consistently belong to the same community across the levels. The resulting community structures also suggest that organizational coalitions in Finnish climate policy, to the extent that retweet coalitions capture them, matter much less past a certain distance from the "core" of the organizations.

Second, as we show in Figure~\ref{fig:snorms_assignments}B, the number of inferred communities differs across organization levels. Notably, we do not find block structure for organization side accounts (indicated with the gray bar), so the solution is one block. This may have been caused by the sparsity of the organization side graph. Among the other levels, we see that, as observed, individual side accounts cluster into a higher number of blocks than do organizational main and individual main accounts, which provides further evidence that the logic of inter-organizational coalitions breaks down as we move to the peripheries of the policy system. However, the difference can also in part be attributed to the overall activity across organization levels, as adjusting the different level graphs to have the same size as the organization main level reduces the number of inferred blocks.

Finally, we use the normalized Reduced Mutual Information (RMI) measure \cite{newman_improved_2020} to evaluate the similarity of our partitions across all bootstrapped networks. The RMI quantifies the amount of information that one obtains from a set of blocks by observing another set of blocks. A value of 1 indicates that the structures are identical, while dissimilar structures tend to exhibit values closer to 0. RMI could also return negative values if the observed associations between partitions are weaker or comparable to what would be expected in a random graph.

Our RMI results are in Figure~\ref{fig:snorms_assignments}C. Here, we do not show results for the organization side accounts because, again, they were too sparse and did not return a block structure. The distributions of the bootstrapped similarity values between different organization levels are all around zero. They are also lower than the bootstrapped similarity values for the levels with themselves. Together these results imply that using different organization levels leads to different inferred communities.

\subsection{Exponential Random Graph Modeling}
Social science research is often interested in explaining the existence or absence of various relationships between actors in a sociopolitical system. These questions are highly relevant in, for example, policy studies \cite{yla2018climate}, where researchers are interested in relationships between actors, such as how they collaborate or share information \cite{leifeld2012information}. They identify factors explaining tie formation in these policy networks, which are then taken to be behavioral tendencies of policy actors that contribute to meaningful relationships in the policy system. For example, information-sharing relations among climate policy actors in the U.S. exhibited tendencies for localized `echo chambers' where actors shared information in dense cliques of similarly minded peers \cite{jasny2015empirical}. With the increased importance of social media space as a forum for policy contestation \cite{theocharis2015conceptualization}, applied researchers have focused more on social media networks \cite[e.g.,][]{li2021organizational,goritz2022international,kotkaniemi2024policy}. As with the analyses of the group structure in social networks, this type of link prediction analyses predominantly do not account for the possibility of their results being specific to a certain organizational level.

To see how differences between organization levels impact conclusions drawn from conventional analyses used in applied research, we conducted a study of the different organizational level networks using the exponential random graph model (ERGM), which is commonly used for this type of policy network analysis \cite[e.g.,][]{leifeld2012information,jasny2015empirical,kotkaniemi2024policy}. The ERGM is a statistical model that allows for inference on tie formation in a network by considering model terms at the vertex (e.g. activity by actor type), dyad (e.g. homophily), and network levels (e.g. triadic closure) \cite{cranmer2011inferential}. More explicitly, in the ERGM, the probability of observing the graph $\boldsymbol{G}$ is given by
\begin{equation}\label{eq:full_ergm}
Pr(\boldsymbol{G},\boldsymbol{\theta})=\kappa^{-1}\exp\{\boldsymbol{\theta}'\boldsymbol{h}(\boldsymbol{G})\}\,,
\end{equation}
where $\boldsymbol{\theta}$ is a vector of coefficients and $\boldsymbol{h}(\boldsymbol{G})$ is a vector of statistics computed over $\boldsymbol{G}$ which constitute the model terms. $\kappa$ is the normalizing constant that transforms the equation into a valid probability density function.

We fit ERGMs for our collapsed graphs at each organizational level and examine how fitted coefficient estimates differ. In our model, we included a sectoral homophily term and a triadic closure term, which are commonly analyzed factors in policy studies \cite{metz2023policy,jasny2015empirical}, while controlling for activity levels by the organization's sector. Sectoral homophily, or the tendency for edges to be formed between organizations from the same sector (e.g. government, business, and civil society), is important to policy studies because it reflects shared preferences or affinities for collaboration based on societal roles associated with different sectors. Triadic closure, formally modeled as the likelihood of an edge to be formed on a dyad that shares at least one partner, captures the tendency for relationships to form closed triangles, contributing to dense local clique-like groups. As all our analysis above, we conduct the entire suite of analysis on the set of bootstrapped networks described in Section~\ref{sec:network_construction}.

\begin{figure}[!t]
    \centering
    \includegraphics[width = 1\textwidth]{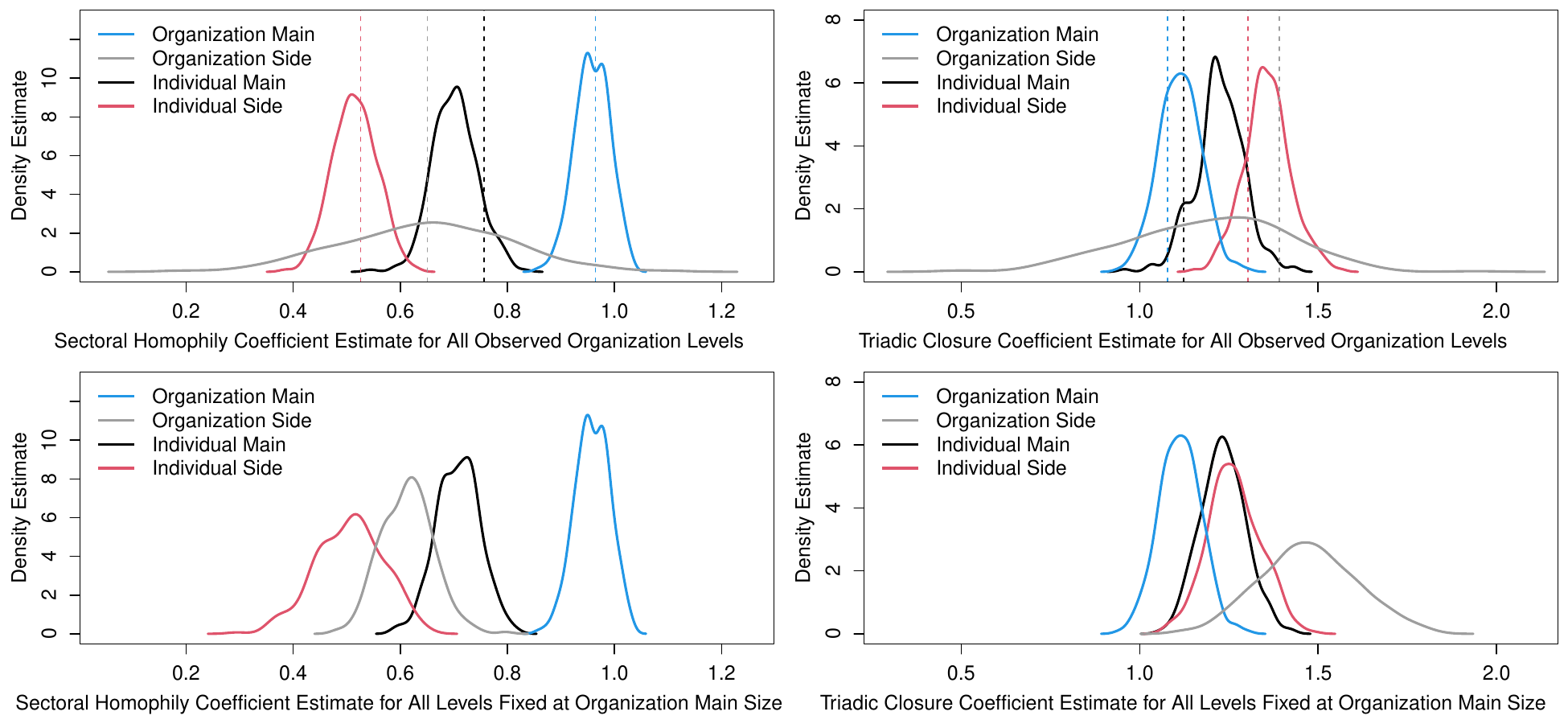}
    \caption{Exponential Random Graph Model Estimates for Homophily and Triadic Closure}
    \label{fig:ergm}
\end{figure}

We report our findings in Figure~\ref{fig:ergm} as bootstrapped distributions of ERGM coefficients. First, we find substantial differences in the sectoral homophily estimates across organization levels. The highest estimates are for the organization main accounts, and decreases as we move to organizationally more peripheral levels, with the individual side accounts havin the lowest estimated tendency for sectoral homophily when interacting within levels. Conditional on endorsements being made, the main official accounts of the organizations are much more likely than accounts from other levels to keep these endorsement to organizations from their own sector, which suggests that they are more cognizant of how policy interests align along sectoral lines. Second, we find smaller but still meaningful differences across organization levels in terms of their tendency for triadic closure. The pattern, however, is the opposite of sectoral homophily, as organization main accounts exhibit the lowest tendency for triadic closure compared to the other levels. Substantively, this means that when interacting within levels, endorsements from organizational main accounts are less likely than those from the other levels to go toward accounts they are already indirectly connected to, which perhaps indicates a relatively low value of indirect relationships when conducting nominally official business.

Finally, similar to the stochastic block model results, the collapsed organization side graph is too sparse to obtain meaningful results in the ERGM analysis. Before fixing the bootstrapped organization side graphs to be the same size as the observed organizational main graph, the estimated coefficient distributions for both sectoral homophily and triadic closure were unreasonably flat and wide, meaning that we cannot confidently infer anything from the fitted model. Accounting for the different organizational levels' graph sizes fixed this problem, as the resulting distributions for the organization side accounts appears to have converged. Fixing the size of the bootstrapped graphs also reduced the differences between organization levels for the triadic closure estimates, but not for the sectoral homophily estimates. While this indicates that some of the observed differences between organization levels are attributable to differences in activity across levels -- which should still be understood as meaningful differences -- there is still a component of unexplained difference between the endorsement tendencies of accounts across organization levels.

\section{Discussion}
Our exploration of social networks at four organization levels in the context of Finnish climate policy revealed significant differences in activity patterns and network structures, with some variations depending on the organization type. These findings have important implications for how researchers define and study organizations on social media.

We found clear distinctions between organizations' official main and side accounts and executive and non-executive personal accounts. When looking at activity, official organization accounts were more pronouncedly active during office hours than the personal accounts. The activity patterns of the official accounts probably reflect the formal employment contracts of the people who have been tasked to manage them.

We also found differences between levels for our network structure measures. The executives' accounts had the highest probability of mutual endorsement, and the highest within-organization densities, neighborhood overlaps, and homophily. This pattern fits with our speculation that executives may have higher incentives to publicly show their alignment with the goals and values of their affiliated organizations than individuals in non-executive roles. The behavior can aslo be affected by the descriptive norms where executives see each other endorsing similar views more than individuals with non-executive roles. Finally,  executives may be closer to each other in other social networks and some possibly also serve on the boards of each other's organizations \cite{bonacich1986interlocking}.

We observed some variation between organization levels across organization types. However, the differences were not large when considering the confidence intervals. The interaction between organization levels and organization types merits further investigation. We also examined the implications of organization level differences on applied research by analysing group sorting (SBM) and tie formation (ERGM). In both cases, the results were sensitive to the organizational levels.

While we expect that our exploratory results are replicable, we recognize that the current data are limited to retweets among organizational actors within the specific context of Finnish climate policy during a specific time period. Climate policy is arguably one of the most politicized topics on social media globally, thereby making it a peculiar case for studying whether the behavior of individuals aligns with that of their respective organizations. Strong politicization could either manifest as a strict top-down imposition of organizational norms or as a greater diversity of opinions among the affiliates of an organization. According to our results, in Finland, which, along with other North European countries, ranks among the most self-expressive cultures in the world \cite{inglehart2005modernization}, the latter option appears more likely. Yet, individual attitudes toward climate policy are often deeply rooted in persistent beliefs, strong emotions, and identification with social groups \cite{hornung2022identities}, which could weaken the effect of loosely enforced top-down organizational norms.

In conclusion, our findings underscore the importance of considering multiple organizational levels when studying social media. Future research should try to replicate the results with different data. It could also study the effects of organization types in more detail and extend the analysis to communication content. As social media continues to provide unprecedented communication opportunities for organizations across all levels, understanding their different characteristics is important for accurate analysis and interpretation of organizational behavior online.

\bibliographystyle{ACM-Reference-Format}
\bibliography{norm}


\begin{thebibliography}{68}


\ifx \showCODEN    \undefined \def \showCODEN     #1{\unskip}     \fi
\ifx \showDOI      \undefined \def \showDOI       #1{#1}\fi
\ifx \showISBNx    \undefined \def \showISBNx     #1{\unskip}     \fi
\ifx \showISBNxiii \undefined \def \showISBNxiii  #1{\unskip}     \fi
\ifx \showISSN     \undefined \def \showISSN      #1{\unskip}     \fi
\ifx \showLCCN     \undefined \def \showLCCN      #1{\unskip}     \fi
\ifx \shownote     \undefined \def \shownote      #1{#1}          \fi
\ifx \showarticletitle \undefined \def \showarticletitle #1{#1}   \fi
\ifx \showURL      \undefined \def \showURL       {\relax}        \fi
\providecommand\bibfield[2]{#2}
\providecommand\bibinfo[2]{#2}
\providecommand\natexlab[1]{#1}
\providecommand\showeprint[2][]{arXiv:#2}

\bibitem[Agerstr{\"o}m et~al\mbox{.}(2016)]%
        {agerstrom2016UsingDescriptiveSocial}
\bibfield{author}{\bibinfo{person}{Jens Agerstr{\"o}m},
  \bibinfo{person}{Rickard Carlsson}, \bibinfo{person}{Linda Nicklasson}, {and}
  \bibinfo{person}{Linda Guntell}.} \bibinfo{year}{2016}\natexlab{}.
\newblock \showarticletitle{Using Descriptive Social Norms to Increase
  Charitable Giving: {{The}} Power of Local Norms}.
\newblock \bibinfo{journal}{\emph{Journal of Economic Psychology}}
  \bibinfo{volume}{52} (\bibinfo{date}{feb} \bibinfo{year}{2016}),
  \bibinfo{pages}{147--153}.
\newblock
\showISSN{0167-4870}
\urldef\tempurl%
\url{https://doi.org/10.1016/j.joep.2015.12.007}
\showDOI{\tempurl}


\bibitem[Bonacich and Roy(2018)]%
        {bonacich1986interlocking}
\bibfield{author}{\bibinfo{person}{Phillip Bonacich} {and}
  \bibinfo{person}{William~G Roy}.} \bibinfo{year}{2018}\natexlab{}.
\newblock \showarticletitle{Centrality, dominance, and interorganizational
  power in a network structure: Interlocking directorates among American
  railroads, 1886–1905}.
\newblock \bibinfo{journal}{\emph{Journal of Mathematical Sociology}}
  \bibinfo{volume}{12}, \bibinfo{number}{2} (\bibinfo{year}{2018}),
  \bibinfo{pages}{127--135}.
\newblock


\bibitem[Chen et~al\mbox{.}(2024)]%
        {chen2024climate}
\bibfield{author}{\bibinfo{person}{Ted Hsuan~Yun Chen}, \bibinfo{person}{Arttu
  Malkam\"aki}, \bibinfo{person}{Ali Faqeeh}, \bibinfo{person}{Esa Palosaari},
  \bibinfo{person}{Anniina Kotkaniemi}, \bibinfo{person}{Laura Funke},
  \bibinfo{person}{C\'ait Gleeson}, \bibinfo{person}{James Goodman},
  \bibinfo{person}{Antti Gronow}, \bibinfo{person}{Marlene Kammerer},
  \bibinfo{person}{Myanna Lahsen}, \bibinfo{person}{Alexandre Marques},
  \bibinfo{person}{Petr Ocelik}, \bibinfo{person}{Shivangi Seth},
  \bibinfo{person}{Mark Stoddart}, \bibinfo{person}{Pradip Swarnakar},
  \bibinfo{person}{Matthew Trull}, \bibinfo{person}{Paul Wagner},
  \bibinfo{person}{Yixi Yang}, \bibinfo{person}{Mikko Kivel\"a}, {and}
  \bibinfo{person}{Tuomas Yl\"a-Anttila}.} \bibinfo{year}{2024}\natexlab{}.
\newblock \bibinfo{title}{Climate Policy Elites' Twitter Interactions across
  Nine Countries.}
\newblock \bibinfo{howpublished}{Working Paper}.
\newblock


\bibitem[Chen et~al\mbox{.}(2021)]%
        {chen2021polarization}
\bibfield{author}{\bibinfo{person}{Ted Hsuan~Yun Chen}, \bibinfo{person}{Ali
  Salloum}, \bibinfo{person}{Antti Gronow}, \bibinfo{person}{Tuomas
  Yl{\"a}-Anttila}, {and} \bibinfo{person}{Mikko Kivel{\"a}}.}
  \bibinfo{year}{2021}\natexlab{}.
\newblock \showarticletitle{Polarization of climate politics results from
  partisan sorting: Evidence from Finnish Twittersphere}.
\newblock \bibinfo{journal}{\emph{Global Environmental Change}}
  \bibinfo{volume}{71} (\bibinfo{year}{2021}), \bibinfo{pages}{102348}.
\newblock


\bibitem[Cheng et~al\mbox{.}(2021)]%
        {cheng2021advocacy}
\bibfield{author}{\bibinfo{person}{Huimin Cheng}, \bibinfo{person}{Ye Wang},
  \bibinfo{person}{Ma Ping}, {and} \bibinfo{person}{Amanda Murdie}.}
  \bibinfo{year}{2021}\natexlab{}.
\newblock \showarticletitle{Communities and Brokers: How the Transnational
  Advocacy Network Simultaneously Provides Social Power and Exacerbates Global
  Inequalities}.
\newblock \bibinfo{journal}{\emph{International Studies Quarterly}}
  \bibinfo{volume}{65}, \bibinfo{number}{3} (\bibinfo{year}{2021}),
  \bibinfo{pages}{724--738}.
\newblock


\bibitem[Coscia and Neffke(2017)]%
        {coscia2017noisy}
\bibfield{author}{\bibinfo{person}{Michele Coscia} {and} \bibinfo{person}{Frank
  Neffke}.} \bibinfo{year}{2017}\natexlab{}.
\newblock \showarticletitle{Network Backboning with Noisy Data}.
\newblock \bibinfo{journal}{\emph{IEEE International Conference on Data
  Engineering}}  \bibinfo{volume}{33} (\bibinfo{year}{2017}),
  \bibinfo{pages}{425--436}.
\newblock


\bibitem[Cranmer and Desmarais(2011)]%
        {cranmer2011inferential}
\bibfield{author}{\bibinfo{person}{Skyler~J Cranmer} {and}
  \bibinfo{person}{Bruce~A Desmarais}.} \bibinfo{year}{2011}\natexlab{}.
\newblock \showarticletitle{Inferential network analysis with exponential
  random graph models}.
\newblock \bibinfo{journal}{\emph{Political analysis}} \bibinfo{volume}{19},
  \bibinfo{number}{1} (\bibinfo{year}{2011}), \bibinfo{pages}{66--86}.
\newblock


\bibitem[De~Choudhury and Counts(2013)]%
        {dechoudhury2013UnderstandingAffectWorkplace}
\bibfield{author}{\bibinfo{person}{Munmun De~Choudhury} {and}
  \bibinfo{person}{Scott Counts}.} \bibinfo{year}{helmikuu 23,
  2013}\natexlab{}.
\newblock \showarticletitle{Understanding Affect in the Workplace via Social
  Media}. In \bibinfo{booktitle}{\emph{Proceedings of the 2013 Conference on
  {{Computer}} Supported Cooperative Work}} \emph{(\bibinfo{series}{{{CSCW}}
  '13})}. \bibinfo{publisher}{{Association for Computing Machinery}},
  \bibinfo{address}{{New York, NY, USA}}, \bibinfo{pages}{303--316}.
\newblock
\showISBNx{978-1-4503-1331-5}
\urldef\tempurl%
\url{https://doi.org/10.1145/2441776.2441812}
\showDOI{\tempurl}


\bibitem[Durkin~Richer et~al\mbox{.}(2024)]%
        {durkinricher2024WhatKnowSupreme}
\bibfield{author}{\bibinfo{person}{Alanna Durkin~Richer}, \bibinfo{person}{Eric
  Tucker}, {and} \bibinfo{person}{Michael Kunzelman}.}
  \bibinfo{year}{2024}\natexlab{}.
\newblock \bibinfo{title}{What to Know about the {{Supreme Court}} Immunity
  Ruling in {{Trump}}'s 2020 Election Interference Case}.
\newblock
\newblock
\urldef\tempurl%
\url{https://apnews.com/article/trump-immunity-supreme-court-capitol-riot-trial-72ec35de776315183e1db561257cb108}
\showURL{%
\tempurl}


\bibitem[Falkenberg et~al\mbox{.}(2022)]%
        {falkenberg2022growing}
\bibfield{author}{\bibinfo{person}{Max Falkenberg}, \bibinfo{person}{Alessandro
  Galeazzi}, \bibinfo{person}{Maddalena Torricelli},
  \bibinfo{person}{Niccol{\`o} Di~Marco}, \bibinfo{person}{Francesca Larosa},
  \bibinfo{person}{Madalina Sas}, \bibinfo{person}{Amin Mekacher},
  \bibinfo{person}{Warren Pearce}, \bibinfo{person}{Fabiana Zollo},
  \bibinfo{person}{Walter Quattrociocchi}, {et~al\mbox{.}}}
  \bibinfo{year}{2022}\natexlab{}.
\newblock \showarticletitle{Growing polarization around climate change on
  social media}.
\newblock \bibinfo{journal}{\emph{Nature Climate Change}} \bibinfo{volume}{12},
  \bibinfo{number}{12} (\bibinfo{year}{2022}), \bibinfo{pages}{1114--1121}.
\newblock


\bibitem[Figenschou and Fredheim(2020)]%
        {figenschou2020interest}
\bibfield{author}{\bibinfo{person}{Tine~Ustad Figenschou} {and}
  \bibinfo{person}{Nanna~Alida Fredheim}.} \bibinfo{year}{2020}\natexlab{}.
\newblock \showarticletitle{Interest groups on social media: Four forms of
  networked advocacy}.
\newblock \bibinfo{journal}{\emph{Journal of Public Affairs}}
  \bibinfo{volume}{20}, \bibinfo{number}{2} (\bibinfo{year}{2020}),
  \bibinfo{pages}{e2012}.
\newblock


\bibitem[Fisher et~al\mbox{.}(2022)]%
        {fisher2022climate}
\bibfield{author}{\bibinfo{person}{Stephen~D Fisher}, \bibinfo{person}{John
  Kenny}, \bibinfo{person}{Wouter Poortinga}, \bibinfo{person}{Gisela Böhm},
  {and} \bibinfo{person}{Linda Steg}.} \bibinfo{year}{2022}\natexlab{}.
\newblock \showarticletitle{The politicisation of climate change attitudes in
  Europe}.
\newblock \bibinfo{journal}{\emph{Electoral Studies}}  \bibinfo{volume}{79}
  (\bibinfo{year}{2022}), \bibinfo{pages}{102499}.
\newblock


\bibitem[Frandsen and Johansen(2011)]%
        {frandsen2011rhetoric}
\bibfield{author}{\bibinfo{person}{Finn Frandsen} {and} \bibinfo{person}{Winni
  Johansen}.} \bibinfo{year}{2011}\natexlab{}.
\newblock \showarticletitle{Rhetoric, Climate Change, and Corporate Identity
  Management}.
\newblock \bibinfo{journal}{\emph{Management Communication Quarterly}}
  \bibinfo{volume}{25}, \bibinfo{number}{3} (\bibinfo{year}{2011}),
  \bibinfo{pages}{511--530}.
\newblock


\bibitem[Gandica et~al\mbox{.}(2016)]%
        {gandica2016OriginBurstinessHuman}
\bibfield{author}{\bibinfo{person}{Yerali Gandica}, \bibinfo{person}{Joao
  Carvalho}, \bibinfo{person}{Fernando Sampaio~Dos Aidos},
  \bibinfo{person}{Renaud Lambiotte}, {and} \bibinfo{person}{Timoteo
  Carletti}.} \bibinfo{year}{2016}\natexlab{}.
\newblock \bibinfo{title}{On the Origin of Burstiness in Human Behavior:
  {{The}} Wikipedia Edits Case}.
\newblock
\newblock
\urldef\tempurl%
\url{https://doi.org/10.48550/arXiv.1601.00864}
\showDOI{\tempurl}
\showeprint[arxiv]{1601.00864}~[physics]


\bibitem[Garimella et~al\mbox{.}(2018)]%
        {garimella2018quantifying}
\bibfield{author}{\bibinfo{person}{Kiran Garimella}, \bibinfo{person}{Gianmarco
  De~Francisci Morales}, \bibinfo{person}{Aristides Gionis}, {and}
  \bibinfo{person}{Michael Mathioudakis}.} \bibinfo{year}{2018}\natexlab{}.
\newblock \showarticletitle{Quantifying controversy on social media}.
\newblock \bibinfo{journal}{\emph{ACM Transactions on Social Computing}}
  \bibinfo{volume}{1}, \bibinfo{number}{1} (\bibinfo{year}{2018}),
  \bibinfo{pages}{1--27}.
\newblock


\bibitem[Garimella et~al\mbox{.}(2016)]%
        {garimella2016quote}
\bibfield{author}{\bibinfo{person}{Kiran Garimella}, \bibinfo{person}{Ingmar
  Weber}, {and} \bibinfo{person}{Munmun De~Choudhury}.}
  \bibinfo{year}{2016}\natexlab{}.
\newblock \showarticletitle{Quote RTs on Twitter: Usage of the new feature for
  political discourse}. In \bibinfo{booktitle}{\emph{Proceedings of the 8th ACM
  Conference on Web Science}}. \bibinfo{pages}{200--204}.
\newblock


\bibitem[Goh and Barab{\'a}si(2008)]%
        {goh2008burstiness}
\bibfield{author}{\bibinfo{person}{K-I Goh} {and} \bibinfo{person}{A-L
  Barab{\'a}si}.} \bibinfo{year}{2008}\natexlab{}.
\newblock \showarticletitle{Burstiness and memory in complex systems}.
\newblock \bibinfo{journal}{\emph{Europhysics Letters}} \bibinfo{volume}{81},
  \bibinfo{number}{4} (\bibinfo{year}{2008}), \bibinfo{pages}{48002}.
\newblock


\bibitem[Goritz et~al\mbox{.}(2022)]%
        {goritz2022international}
\bibfield{author}{\bibinfo{person}{Alexandra Goritz}, \bibinfo{person}{Johannes
  Schuster}, \bibinfo{person}{Helge J{\"o}rgens}, {and} \bibinfo{person}{Nina
  Kolleck}.} \bibinfo{year}{2022}\natexlab{}.
\newblock \showarticletitle{International public administrations on Twitter: A
  comparison of digital authority in global climate policy}.
\newblock \bibinfo{journal}{\emph{Journal of Comparative Policy Analysis:
  Research and Practice}} \bibinfo{volume}{24}, \bibinfo{number}{3}
  (\bibinfo{year}{2022}), \bibinfo{pages}{271--295}.
\newblock


\bibitem[Granovetter(1973)]%
        {granovetter1973strength}
\bibfield{author}{\bibinfo{person}{Mark~S Granovetter}.}
  \bibinfo{year}{1973}\natexlab{}.
\newblock \showarticletitle{The strength of weak ties}.
\newblock \bibinfo{journal}{\emph{American journal of sociology}}
  \bibinfo{volume}{78}, \bibinfo{number}{6} (\bibinfo{year}{1973}),
  \bibinfo{pages}{1360--1380}.
\newblock


\bibitem[Gronow and Malkamäki(2024)]%
        {gronow2024partisan}
\bibfield{author}{\bibinfo{person}{Antti Gronow} {and} \bibinfo{person}{Arttu
  Malkamäki}.} \bibinfo{year}{2024}\natexlab{}.
\newblock \showarticletitle{Political polarisation in turbulent times: Tracking
  polarisation trends and partisan news link sharing on Finnish Twitter,
  2015-2023}.
\newblock \bibinfo{journal}{\emph{arXiv:2403.03842}} (\bibinfo{year}{2024}).
\newblock


\bibitem[Hayes and Scott(2018)]%
        {hayes2018multiplex}
\bibfield{author}{\bibinfo{person}{Adam~L Hayes} {and} \bibinfo{person}{Tyler~A
  Scott}.} \bibinfo{year}{2018}\natexlab{}.
\newblock \showarticletitle{Multiplex network analysis for complex governance
  systems using surveys and online behavior}.
\newblock \bibinfo{journal}{\emph{Policy Studies Journal}}
  \bibinfo{volume}{46}, \bibinfo{number}{2} (\bibinfo{year}{2018}),
  \bibinfo{pages}{327--353}.
\newblock


\bibitem[Hemphill et~al\mbox{.}(2013)]%
        {hemphill2013}
\bibfield{author}{\bibinfo{person}{Libby Hemphill}, \bibinfo{person}{Jahna
  Otterbacher}, {and} \bibinfo{person}{Matthew Shapiro}.}
  \bibinfo{year}{2013}\natexlab{}.
\newblock \showarticletitle{What's congress doing on twitter?}. In
  \bibinfo{booktitle}{\emph{Proceedings of the 2013 Conference on Computer
  Supported Cooperative Work}} (San Antonio, Texas, USA)
  \emph{(\bibinfo{series}{CSCW '13})}. \bibinfo{publisher}{Association for
  Computing Machinery}, \bibinfo{address}{New York, NY, USA},
  \bibinfo{pages}{877–886}.
\newblock
\showISBNx{9781450313315}
\urldef\tempurl%
\url{https://doi.org/10.1145/2441776.2441876}
\showDOI{\tempurl}


\bibitem[Holland et~al\mbox{.}(1983)]%
        {holland1983stochastic}
\bibfield{author}{\bibinfo{person}{Paul~W Holland},
  \bibinfo{person}{Kathryn~Blackmond Laskey}, {and} \bibinfo{person}{Samuel
  Leinhardt}.} \bibinfo{year}{1983}\natexlab{}.
\newblock \showarticletitle{Stochastic blockmodels: First steps}.
\newblock \bibinfo{journal}{\emph{Social networks}} \bibinfo{volume}{5},
  \bibinfo{number}{2} (\bibinfo{year}{1983}), \bibinfo{pages}{109--137}.
\newblock


\bibitem[Hornung(2022)]%
        {hornung2022identities}
\bibfield{author}{\bibinfo{person}{Johanna Hornung}.}
  \bibinfo{year}{2022}\natexlab{}.
\newblock \showarticletitle{Social identities in climate action}.
\newblock \bibinfo{journal}{\emph{Climate Action}}  \bibinfo{volume}{1}
  (\bibinfo{year}{2022}), \bibinfo{pages}{4}.
\newblock


\bibitem[Inglehart and Welzel(2005)]%
        {inglehart2005modernization}
\bibfield{author}{\bibinfo{person}{Ronald Inglehart} {and}
  \bibinfo{person}{Christian Welzel}.} \bibinfo{year}{2005}\natexlab{}.
\newblock \bibinfo{booktitle}{\emph{Modernization, Cultural Change, and
  Democracy: The Human Development Sequence}}.
\newblock \bibinfo{publisher}{Cambridge University Press}.
\newblock


\bibitem[Jacobs and Watts(2021)]%
        {jacobs2021LargeScaleComparativeStudy}
\bibfield{author}{\bibinfo{person}{Abigail~Z. Jacobs} {and}
  \bibinfo{person}{Duncan~J. Watts}.} \bibinfo{year}{2021}\natexlab{}.
\newblock \showarticletitle{A {{Large-Scale Comparative Study}} of {{Informal
  Social Networks}} in {{Firms}}}.
\newblock \bibinfo{journal}{\emph{Management Science}} \bibinfo{volume}{67},
  \bibinfo{number}{9} (\bibinfo{date}{Sept.} \bibinfo{year}{2021}),
  \bibinfo{pages}{5489--5509}.
\newblock
\showISSN{0025-1909}
\urldef\tempurl%
\url{https://doi.org/10.1287/mnsc.2021.3997}
\showDOI{\tempurl}


\bibitem[Jasny et~al\mbox{.}(2015)]%
        {jasny2015empirical}
\bibfield{author}{\bibinfo{person}{Lorien Jasny}, \bibinfo{person}{Joseph
  Waggle}, {and} \bibinfo{person}{Dana~R Fisher}.}
  \bibinfo{year}{2015}\natexlab{}.
\newblock \showarticletitle{An empirical examination of echo chambers in US
  climate policy networks}.
\newblock \bibinfo{journal}{\emph{Nature Climate Change}} \bibinfo{volume}{5},
  \bibinfo{number}{8} (\bibinfo{year}{2015}), \bibinfo{pages}{782--786}.
\newblock


\bibitem[Karrer and Newman(2011)]%
        {karrer_stochastic_2011}
\bibfield{author}{\bibinfo{person}{Brian Karrer} {and}
  \bibinfo{person}{M.~E.~J. Newman}.} \bibinfo{year}{2011}\natexlab{}.
\newblock \showarticletitle{Stochastic blockmodels and community structure in
  networks}.
\newblock \bibinfo{journal}{\emph{Physical Review E}} \bibinfo{volume}{83},
  \bibinfo{number}{1} (\bibinfo{year}{2011}), \bibinfo{pages}{016107}.
\newblock
\urldef\tempurl%
\url{https://doi.org/10.1103/PhysRevE.83.016107}
\showDOI{\tempurl}


\bibitem[Karsai et~al\mbox{.}(2018)]%
        {karsai2018bursty}
\bibfield{author}{\bibinfo{person}{M{\'a}rton Karsai},
  \bibinfo{person}{Hang-Hyun Jo}, \bibinfo{person}{Kimmo Kaski},
  {et~al\mbox{.}}} \bibinfo{year}{2018}\natexlab{}.
\newblock \bibinfo{booktitle}{\emph{Bursty human dynamics}}.
\newblock \bibinfo{publisher}{Springer}.
\newblock


\bibitem[Kim(2023)]%
        {kim2023MuskVowsPay}
\bibfield{author}{\bibinfo{person}{Juliana Kim}.}
  \bibinfo{year}{2023}\natexlab{}.
\newblock \showarticletitle{Musk Vows to Pay Legal Costs for Users Who Get in
  Trouble at Work for Their Tweets}.
\newblock \bibinfo{journal}{\emph{NPR}} (\bibinfo{date}{aug}
  \bibinfo{year}{2023}).
\newblock
\urldef\tempurl%
\url{https://www.npr.org/2023/08/06/1192394226/musk-vows-to-pay-legal-costs-for-users-who-get-in-trouble-at-work-for-their-twee}
\showURL{%
\tempurl}


\bibitem[Kotkaniemi et~al\mbox{.}(2024)]%
        {kotkaniemi2024policy}
\bibfield{author}{\bibinfo{person}{Anniina Kotkaniemi}, \bibinfo{person}{Tuomas
  Yl{\"a}-Anttila}, {and} \bibinfo{person}{Ted Hsuan~Yun Chen}.}
  \bibinfo{year}{2024}\natexlab{}.
\newblock \showarticletitle{Policy influence and influencers online and off}.
\newblock \bibinfo{journal}{\emph{Policy Studies Journal}}
  (\bibinfo{year}{2024}).
\newblock


\bibitem[Lapinski and Rimal(2005)]%
        {lapinski2005ExplicationSocialNorms}
\bibfield{author}{\bibinfo{person}{Maria~Knight Lapinski} {and}
  \bibinfo{person}{Rajiv~N. Rimal}.} \bibinfo{year}{2005}\natexlab{}.
\newblock \showarticletitle{An {{Explication}} of {{Social Norms}}}.
\newblock \bibinfo{journal}{\emph{Communication Theory}} \bibinfo{volume}{15},
  \bibinfo{number}{2} (\bibinfo{year}{2005}), \bibinfo{pages}{127--147}.
\newblock
\showISSN{1468-2885}
\urldef\tempurl%
\url{https://doi.org/10.1111/j.1468-2885.2005.tb00329.x}
\showDOI{\tempurl}


\bibitem[Leifeld and Schneider(2012)]%
        {leifeld2012information}
\bibfield{author}{\bibinfo{person}{Philip Leifeld} {and}
  \bibinfo{person}{Volker Schneider}.} \bibinfo{year}{2012}\natexlab{}.
\newblock \showarticletitle{Information exchange in policy networks}.
\newblock \bibinfo{journal}{\emph{American Journal of Political Science}}
  \bibinfo{volume}{56}, \bibinfo{number}{3} (\bibinfo{year}{2012}),
  \bibinfo{pages}{731--744}.
\newblock


\bibitem[Li et~al\mbox{.}(2021)]%
        {li2021organizational}
\bibfield{author}{\bibinfo{person}{Yiqi Li}, \bibinfo{person}{Jieun Shin},
  \bibinfo{person}{JIngyi Sun}, \bibinfo{person}{Hye~Min Kim},
  \bibinfo{person}{Yan Qu}, {and} \bibinfo{person}{Aimei Yang}.}
  \bibinfo{year}{2021}\natexlab{}.
\newblock \showarticletitle{Organizational sensemaking in tough times: The
  ecology of NGOs’ COVID-19 issue discourse communities on social media}.
\newblock \bibinfo{journal}{\emph{Computers in human behavior}}
  \bibinfo{volume}{122} (\bibinfo{year}{2021}), \bibinfo{pages}{106838}.
\newblock


\bibitem[Lorrain and White(1971)]%
        {lorrain1971structural}
\bibfield{author}{\bibinfo{person}{Francois Lorrain} {and}
  \bibinfo{person}{Harrison~C White}.} \bibinfo{year}{1971}\natexlab{}.
\newblock \showarticletitle{Structural equivalence of individuals in social
  networks}.
\newblock \bibinfo{journal}{\emph{The Journal of mathematical sociology}}
  \bibinfo{volume}{1}, \bibinfo{number}{1} (\bibinfo{year}{1971}),
  \bibinfo{pages}{49--80}.
\newblock


\bibitem[Lovejoy and Saxton(2012)]%
        {lovejoy2012}
\bibfield{author}{\bibinfo{person}{Kristen Lovejoy} {and}
  \bibinfo{person}{Gregory~D. Saxton}.} \bibinfo{year}{2012}\natexlab{}.
\newblock \showarticletitle{{Information, Community, and Action: How Nonprofit
  Organizations Use Social Media*}}.
\newblock \bibinfo{journal}{\emph{Journal of Computer-Mediated Communication}}
  \bibinfo{volume}{17}, \bibinfo{number}{3} (\bibinfo{date}{04}
  \bibinfo{year}{2012}), \bibinfo{pages}{337--353}.
\newblock
\showISSN{1083-6101}
\urldef\tempurl%
\url{https://doi.org/10.1111/j.1083-6101.2012.01576.x}
\showDOI{\tempurl}
\showeprint{https://academic.oup.com/jcmc/article-pdf/17/3/337/19492658/jjcmcom0337.pdf}


\bibitem[Malkamäki et~al\mbox{.}(2023)]%
        {malkamaki2023complex}
\bibfield{author}{\bibinfo{person}{Arttu Malkamäki}, \bibinfo{person}{Ted
  Hsuan~Yun Chen}, \bibinfo{person}{Antti Gronow}, \bibinfo{person}{Mikko
  Kivelä}, \bibinfo{person}{Juho Vesa}, {and} \bibinfo{person}{Tuomas
  Ylä-Anttila}.} \bibinfo{year}{2023}\natexlab{}.
\newblock \showarticletitle{Complex coalitions: political alliances across
  relational contexts}.
\newblock \bibinfo{journal}{\emph{arXiv:2308.14422}} (\bibinfo{year}{2023}).
\newblock


\bibitem[Mattie and Onnela(2017)]%
        {mattie2017generalizations}
\bibfield{author}{\bibinfo{person}{Heather Mattie} {and}
  \bibinfo{person}{Jukka-Pekka Onnela}.} \bibinfo{year}{2017}\natexlab{}.
\newblock \showarticletitle{Generalizations of edge overlap to weighted and
  directed networks}.
\newblock \bibinfo{journal}{\emph{arXiv:1712.07110}} (\bibinfo{year}{2017}).
\newblock


\bibitem[Metaxas et~al\mbox{.}(2015)]%
        {metaxas2015retweets}
\bibfield{author}{\bibinfo{person}{Panagiotis Metaxas}, \bibinfo{person}{Eni
  Mustafaraj}, \bibinfo{person}{Kily Wong}, \bibinfo{person}{Laura Zeng},
  \bibinfo{person}{Megan O'Keefe}, {and} \bibinfo{person}{Samantha Finn}.}
  \bibinfo{year}{2015}\natexlab{}.
\newblock \showarticletitle{What do retweets indicate? Results from user survey
  and meta-review of research}. In \bibinfo{booktitle}{\emph{Proceedings of the
  international AAAI conference on web and social media}},
  Vol.~\bibinfo{volume}{9}. \bibinfo{pages}{658--661}.
\newblock


\bibitem[Metz and Brandenberger(2023)]%
        {metz2023policy}
\bibfield{author}{\bibinfo{person}{Florence Metz} {and}
  \bibinfo{person}{Laurence Brandenberger}.} \bibinfo{year}{2023}\natexlab{}.
\newblock \showarticletitle{Policy networks across political systems}.
\newblock \bibinfo{journal}{\emph{American journal of political science}}
  \bibinfo{volume}{67}, \bibinfo{number}{3} (\bibinfo{year}{2023}),
  \bibinfo{pages}{569--586}.
\newblock


\bibitem[Neal(2022)]%
        {neal2022backbone}
\bibfield{author}{\bibinfo{person}{Zachary Neal}.}
  \bibinfo{year}{2022}\natexlab{}.
\newblock \showarticletitle{Political polarisation in turbulent times: Tracking
  polarisation trends and partisan news link sharing on Finnish Twitter,
  2015-2023}.
\newblock \bibinfo{journal}{\emph{PLoS ONE}} \bibinfo{volume}{17},
  \bibinfo{number}{5} (\bibinfo{year}{2022}), \bibinfo{pages}{e0269137}.
\newblock


\bibitem[Newman et~al\mbox{.}(2020)]%
        {newman_improved_2020}
\bibfield{author}{\bibinfo{person}{M.~E.~J. Newman}, \bibinfo{person}{George~T.
  Cantwell}, {and} \bibinfo{person}{Jean-Gabriel Young}.}
  \bibinfo{year}{2020}\natexlab{}.
\newblock \showarticletitle{Improved mutual information measure for clustering,
  classification, and community detection}.
\newblock \bibinfo{journal}{\emph{Physical Review E}} \bibinfo{volume}{101},
  \bibinfo{number}{4} (\bibinfo{year}{2020}), \bibinfo{pages}{042304}.
\newblock
\urldef\tempurl%
\url{https://doi.org/10.1103/PhysRevE.101.042304}
\showDOI{\tempurl}


\bibitem[Oberg and Walgenbach(2008)]%
        {oberg2008HierarchicalStructuresCommunication}
\bibfield{author}{\bibinfo{person}{Achim Oberg} {and} \bibinfo{person}{Peter
  Walgenbach}.} \bibinfo{year}{2008}\natexlab{}.
\newblock \showarticletitle{Hierarchical Structures of Communication in a
  Network Organization}.
\newblock \bibinfo{journal}{\emph{Scandinavian Journal of Management}}
  \bibinfo{volume}{24}, \bibinfo{number}{3} (\bibinfo{date}{Sept.}
  \bibinfo{year}{2008}), \bibinfo{pages}{183--198}.
\newblock
\showISSN{0956-5221}
\urldef\tempurl%
\url{https://doi.org/10.1016/j.scaman.2008.03.011}
\showDOI{\tempurl}


\bibitem[Onnela et~al\mbox{.}(2007)]%
        {onnela2007analysis}
\bibfield{author}{\bibinfo{person}{Jukka-Pekka Onnela}, \bibinfo{person}{Jari
  Saram{\"a}ki}, \bibinfo{person}{J{\"o}rkki Hyv{\"o}nen},
  \bibinfo{person}{G{\'a}bor Szab{\'o}}, \bibinfo{person}{M~Argollo
  De~Menezes}, \bibinfo{person}{Kimmo Kaski},
  \bibinfo{person}{Albert-L{\'a}szl{\'o} Barab{\'a}si}, {and}
  \bibinfo{person}{J{\'a}nos Kert{\'e}sz}.} \bibinfo{year}{2007}\natexlab{}.
\newblock \showarticletitle{Analysis of a large-scale weighted network of
  one-to-one human communication}.
\newblock \bibinfo{journal}{\emph{New journal of physics}} \bibinfo{volume}{9},
  \bibinfo{number}{6} (\bibinfo{year}{2007}), \bibinfo{pages}{179}.
\newblock


\bibitem[Orlikowski(1992)]%
        {orlikowski1992DualityTechnologyRethinking}
\bibfield{author}{\bibinfo{person}{Wanda~J. Orlikowski}.}
  \bibinfo{year}{1992}\natexlab{}.
\newblock \showarticletitle{The {{Duality}} of {{Technology}}: {{Rethinking}}
  the {{Concept}} of {{Technology}} in {{Organizations}}}.
\newblock \bibinfo{journal}{\emph{Organization Science}} \bibinfo{volume}{3},
  \bibinfo{number}{3} (\bibinfo{date}{Aug.} \bibinfo{year}{1992}),
  \bibinfo{pages}{398--427}.
\newblock
\showISSN{1047-7039, 1526-5455}
\urldef\tempurl%
\url{https://doi.org/10.1287/orsc.3.3.398}
\showDOI{\tempurl}


\bibitem[Orlikowski(2007)]%
        {orlikowski2007SociomaterialPracticesExploring}
\bibfield{author}{\bibinfo{person}{Wanda~J. Orlikowski}.}
  \bibinfo{year}{2007}\natexlab{}.
\newblock \showarticletitle{Sociomaterial {{Practices}}: {{Exploring
  Technology}} at {{Work}}}.
\newblock \bibinfo{journal}{\emph{Organization Studies}} \bibinfo{volume}{28},
  \bibinfo{number}{9} (\bibinfo{date}{Sept.} \bibinfo{year}{2007}),
  \bibinfo{pages}{1435--1448}.
\newblock
\showISSN{0170-8406}
\urldef\tempurl%
\url{https://doi.org/10.1177/0170840607081138}
\showDOI{\tempurl}


\bibitem[Peixoto(2014)]%
        {peixoto_hierarchical_2014}
\bibfield{author}{\bibinfo{person}{Tiago~P. Peixoto}.}
  \bibinfo{year}{2014}\natexlab{}.
\newblock \showarticletitle{Hierarchical Block Structures and High-Resolution
  Model Selection in Large Networks}.
\newblock \bibinfo{journal}{\emph{Physical Review X}} \bibinfo{volume}{4},
  \bibinfo{number}{1} (\bibinfo{year}{2014}), \bibinfo{pages}{011047}.
\newblock
\urldef\tempurl%
\url{https://doi.org/10.1103/PhysRevX.4.011047}
\showDOI{\tempurl}


\bibitem[Peixoto(2020)]%
        {peixoto_merge_2020}
\bibfield{author}{\bibinfo{person}{Tiago~P. Peixoto}.}
  \bibinfo{year}{2020}\natexlab{}.
\newblock \showarticletitle{Merge-split Markov chain Monte Carlo for community
  detection}.
\newblock \bibinfo{journal}{\emph{Physical Review E}} \bibinfo{volume}{102},
  \bibinfo{number}{1} (\bibinfo{year}{2020}), \bibinfo{pages}{012305}.
\newblock
\urldef\tempurl%
\url{https://doi.org/10.1103/PhysRevE.102.012305}
\showDOI{\tempurl}


\bibitem[Porter et~al\mbox{.}(2009)]%
        {porter_communities_2009}
\bibfield{author}{\bibinfo{person}{Mason~A. Porter},
  \bibinfo{person}{Jukka-Pekka Onnela}, {and} \bibinfo{person}{Peter~J.
  Mucha}.} \bibinfo{year}{2009}\natexlab{}.
\newblock \showarticletitle{Communities in {Networks}}.
\newblock \bibinfo{journal}{\emph{Notices of the American Mathematical
  Society}} \bibinfo{volume}{56}, \bibinfo{number}{9} (\bibinfo{year}{2009}),
  \bibinfo{pages}{1082--1097}.
\newblock


\bibitem[Riemer et~al\mbox{.}(2015)]%
        {riemer2015TopBottom}
\bibfield{author}{\bibinfo{person}{Kai Riemer}, \bibinfo{person}{Stefan
  Stieglitz}, {and} \bibinfo{person}{Christian Meske}.}
  \bibinfo{year}{2015}\natexlab{}.
\newblock \showarticletitle{From {{Top}} to {{Bottom}}}.
\newblock \bibinfo{journal}{\emph{Business \& Information Systems Engineering}}
  \bibinfo{volume}{57}, \bibinfo{number}{3} (\bibinfo{date}{June}
  \bibinfo{year}{2015}), \bibinfo{pages}{197--212}.
\newblock
\showISSN{1867-0202}
\urldef\tempurl%
\url{https://doi.org/10.1007/s12599-015-0375-3}
\showDOI{\tempurl}


\bibitem[Rybalko and Seltzer(2010)]%
        {rybalko2010dialogic}
\bibfield{author}{\bibinfo{person}{Svetlana Rybalko} {and}
  \bibinfo{person}{Trent Seltzer}.} \bibinfo{year}{2010}\natexlab{}.
\newblock \showarticletitle{Dialogic communication in 140 characters or less:
  How Fortune 500 companies engage stakeholders using Twitter}.
\newblock \bibinfo{journal}{\emph{Public relations review}}
  \bibinfo{volume}{36}, \bibinfo{number}{4} (\bibinfo{year}{2010}),
  \bibinfo{pages}{336--341}.
\newblock


\bibitem[Salloum et~al\mbox{.}(2022)]%
        {salloum2022separating}
\bibfield{author}{\bibinfo{person}{Ali Salloum}, \bibinfo{person}{Ted Hsuan~Yun
  Chen}, {and} \bibinfo{person}{Mikko Kivel{\"a}}.}
  \bibinfo{year}{2022}\natexlab{}.
\newblock \showarticletitle{Separating polarization from noise: comparison and
  normalization of structural polarization measures}.
\newblock \bibinfo{journal}{\emph{Proceedings of the ACM on human-computer
  interaction}} \bibinfo{volume}{6}, \bibinfo{number}{CSCW1}
  (\bibinfo{year}{2022}), \bibinfo{pages}{1--33}.
\newblock


\bibitem[Schmidt(1994)]%
        {schmidt1994OrganizationCooperativeWork}
\bibfield{author}{\bibinfo{person}{Kjeld Schmidt}.}
  \bibinfo{year}{1994}\natexlab{}.
\newblock \showarticletitle{The Organization of Cooperative Work: Beyond the
  ``{{Leviathan}}'' Conception of the Organization of Cooperative Work}. In
  \bibinfo{booktitle}{\emph{Proceedings of the 1994 {{ACM}} Conference on
  {{Computer}} Supported Cooperative Work}} \emph{(\bibinfo{series}{{{CSCW}}
  '94})}. \bibinfo{publisher}{Association for Computing Machinery},
  \bibinfo{address}{New York, NY, USA}, \bibinfo{pages}{101--112}.
\newblock
\showISBNx{978-0-89791-689-9}
\urldef\tempurl%
\url{https://doi.org/10.1145/192844.192883}
\showDOI{\tempurl}


\bibitem[Smith et~al\mbox{.}(2023)]%
        {smith2023partisanship}
\bibfield{author}{\bibinfo{person}{E.~Keith Smith}, \bibinfo{person}{M.~Julia
  Bognar}, {and} \bibinfo{person}{Adam~P. Mayer}.}
  \bibinfo{year}{2023}\natexlab{}.
\newblock \showarticletitle{Multidimensional partisanship shapes climate policy
  support and behaviours}.
\newblock \bibinfo{journal}{\emph{Nature Climate Change}} \bibinfo{volume}{13},
  \bibinfo{number}{1} (\bibinfo{year}{2023}), \bibinfo{pages}{32--39}.
\newblock


\bibitem[Srivastava and Banaji(2011)]%
        {srivastava2011CultureCognitionCollaborative}
\bibfield{author}{\bibinfo{person}{Sameer~B. Srivastava} {and}
  \bibinfo{person}{Mahzarin~R. Banaji}.} \bibinfo{year}{2011}\natexlab{}.
\newblock \showarticletitle{Culture, {{Cognition}}, and {{Collaborative
  Networks}} in {{Organizations}}}.
\newblock \bibinfo{journal}{\emph{American Sociological Review}}
  \bibinfo{volume}{76}, \bibinfo{number}{2} (\bibinfo{date}{April}
  \bibinfo{year}{2011}), \bibinfo{pages}{207--233}.
\newblock
\showISSN{0003-1224}
\urldef\tempurl%
\url{https://doi.org/10.1177/0003122411399390}
\showDOI{\tempurl}


\bibitem[Stanojevic et~al\mbox{.}(2020)]%
        {stanojevic2020GoodWorkersCrooked}
\bibfield{author}{\bibinfo{person}{Antonia Stanojevic}, \bibinfo{person}{Agnes
  Akkerman}, {and} \bibinfo{person}{Katerina Manevska}.}
  \bibinfo{year}{2020}\natexlab{}.
\newblock \showarticletitle{Good {{Workers}} and {{Crooked Bosses}}: {{The
  Effect}} of {{Voice Suppression}} by {{Supervisors}} on {{Employees}}'
  {{Populist Attitudes}} and {{Voting}}}.
\newblock \bibinfo{journal}{\emph{Political Psychology}} \bibinfo{volume}{41},
  \bibinfo{number}{2} (\bibinfo{year}{2020}), \bibinfo{pages}{363--381}.
\newblock
\showISSN{1467-9221}
\urldef\tempurl%
\url{https://doi.org/10.1111/pops.12619}
\showDOI{\tempurl}


\bibitem[Starbird et~al\mbox{.}(2019)]%
        {starbird2019DisinformationCollaborativeWork}
\bibfield{author}{\bibinfo{person}{Kate Starbird}, \bibinfo{person}{Ahmer
  Arif}, {and} \bibinfo{person}{Tom Wilson}.} \bibinfo{year}{marraskuu 7,
  2019}\natexlab{}.
\newblock \showarticletitle{Disinformation as {{Collaborative Work}}:
  {{Surfacing}} the {{Participatory Nature}} of {{Strategic Information
  Operations}}}.
\newblock \bibinfo{journal}{\emph{Proc. ACM Hum.-Comput. Interact.}}
  \bibinfo{volume}{3}, \bibinfo{number}{CSCW} (\bibinfo{year}{marraskuu 7,
  2019}), \bibinfo{pages}{127:1--127:26}.
\newblock
\urldef\tempurl%
\url{https://doi.org/10.1145/3359229}
\showDOI{\tempurl}


\bibitem[Theocharis(2015)]%
        {theocharis2015conceptualization}
\bibfield{author}{\bibinfo{person}{Yannis Theocharis}.}
  \bibinfo{year}{2015}\natexlab{}.
\newblock \showarticletitle{The conceptualization of digitally networked
  participation}.
\newblock \bibinfo{journal}{\emph{Social Media+ Society}} \bibinfo{volume}{1},
  \bibinfo{number}{2} (\bibinfo{year}{2015}),
  \bibinfo{pages}{2056305115610140}.
\newblock


\bibitem[Ure{\~n}a-Carrion et~al\mbox{.}(2020)]%
        {urena2020estimating}
\bibfield{author}{\bibinfo{person}{Javier Ure{\~n}a-Carrion},
  \bibinfo{person}{Jari Saram{\"a}ki}, {and} \bibinfo{person}{Mikko
  Kivel{\"a}}.} \bibinfo{year}{2020}\natexlab{}.
\newblock \showarticletitle{Estimating tie strength in social networks using
  temporal communication data}.
\newblock \bibinfo{journal}{\emph{EPJ Data Science}} \bibinfo{volume}{9},
  \bibinfo{number}{1} (\bibinfo{year}{2020}), \bibinfo{pages}{37}.
\newblock


\bibitem[Van~der Graaf et~al\mbox{.}(2016)]%
        {van2016weapon}
\bibfield{author}{\bibinfo{person}{Amber Van~der Graaf}, \bibinfo{person}{Simon
  Otjes}, {and} \bibinfo{person}{Anne Rasmussen}.}
  \bibinfo{year}{2016}\natexlab{}.
\newblock \showarticletitle{Weapon of the weak? The social media landscape of
  interest groups}.
\newblock \bibinfo{journal}{\emph{European journal of communication}}
  \bibinfo{volume}{31}, \bibinfo{number}{2} (\bibinfo{year}{2016}),
  \bibinfo{pages}{120--135}.
\newblock


\bibitem[Vu et~al\mbox{.}(2020)]%
        {vu2020leads}
\bibfield{author}{\bibinfo{person}{Hong~Tien Vu}, \bibinfo{person}{Hung~Viet
  Do}, \bibinfo{person}{Hyunjin Seo}, {and} \bibinfo{person}{Yuchen Liu}.}
  \bibinfo{year}{2020}\natexlab{}.
\newblock \showarticletitle{Who leads the conversation on climate change?: A
  study of a global network of NGOs on Twitter}.
\newblock \bibinfo{journal}{\emph{Environmental Communication}}
  \bibinfo{volume}{14}, \bibinfo{number}{4} (\bibinfo{year}{2020}),
  \bibinfo{pages}{450--464}.
\newblock


\bibitem[Wallen and Romulo(2017)]%
        {wallen2017SocialNormsMore}
\bibfield{author}{\bibinfo{person}{Kenneth~E. Wallen} {and}
  \bibinfo{person}{Chelsie~L. Romulo}.} \bibinfo{year}{2017}\natexlab{}.
\newblock \showarticletitle{Social Norms: {{More}} Details, Please}.
\newblock \bibinfo{journal}{\emph{Proceedings of the National Academy of
  Sciences}} \bibinfo{volume}{114}, \bibinfo{number}{27} (\bibinfo{date}{July}
  \bibinfo{year}{2017}).
\newblock
\showISSN{0027-8424, 1091-6490}
\urldef\tempurl%
\url{https://doi.org/10.1073/pnas.1704451114}
\showDOI{\tempurl}


\bibitem[Wallis and Loy(2021)]%
        {wallis2021activism}
\bibfield{author}{\bibinfo{person}{Hannah Wallis} {and}
  \bibinfo{person}{Laura~S. Loy}.} \bibinfo{year}{2021}\natexlab{}.
\newblock \showarticletitle{What drives pro-environmental activism of young
  people? A survey study on the Fridays For Future movement}.
\newblock \bibinfo{journal}{\emph{Journal of Environmental Psychology}}
  \bibinfo{volume}{74} (\bibinfo{year}{2021}), \bibinfo{pages}{101581}.
\newblock
\urldef\tempurl%
\url{https://doi.org/10.1016/j.jenvp.2021.101581}
\showDOI{\tempurl}


\bibitem[Wasserman and Faust(1994)]%
        {wasserman1994social}
\bibfield{author}{\bibinfo{person}{Stanley Wasserman} {and}
  \bibinfo{person}{Katherine Faust}.} \bibinfo{year}{1994}\natexlab{}.
\newblock \bibinfo{booktitle}{\emph{Social Network Analysis: Methods and
  Applications}}.
\newblock \bibinfo{publisher}{Cambridge University Press}.
\newblock


\bibitem[Whitesell et~al\mbox{.}(2023)]%
        {whitesell2023local}
\bibfield{author}{\bibinfo{person}{Anne Whitesell}, \bibinfo{person}{Kevin
  Reuning}, {and} \bibinfo{person}{A~Lee Hannah}.}
  \bibinfo{year}{2023}\natexlab{}.
\newblock \showarticletitle{Local political party presence online}.
\newblock \bibinfo{journal}{\emph{Party Politics}} \bibinfo{volume}{29},
  \bibinfo{number}{1} (\bibinfo{year}{2023}), \bibinfo{pages}{164--175}.
\newblock


\bibitem[Xia et~al\mbox{.}(2021)]%
        {xia2021spread}
\bibfield{author}{\bibinfo{person}{Yan Xia}, \bibinfo{person}{Ted Hsuan~Yun
  Chen}, {and} \bibinfo{person}{Mikko Kivel{\"a}}.}
  \bibinfo{year}{2021}\natexlab{}.
\newblock \showarticletitle{Spread of tweets in climate discussions: A case
  study of the 2019 Nobel Peace Prize announcement}.
\newblock \bibinfo{journal}{\emph{Nordic journal of media studies}}
  \bibinfo{volume}{3}, \bibinfo{number}{1} (\bibinfo{year}{2021}),
  \bibinfo{pages}{96--117}.
\newblock


\bibitem[Yl{\"a}-Anttila et~al\mbox{.}(2018)]%
        {yla2018climate}
\bibfield{author}{\bibinfo{person}{Tuomas Yl{\"a}-Anttila},
  \bibinfo{person}{Antti Gronow}, \bibinfo{person}{Mark~CJ Stoddart},
  \bibinfo{person}{Jeffrey Broadbent}, \bibinfo{person}{Volker Schneider},
  {and} \bibinfo{person}{David~B Tindall}.} \bibinfo{year}{2018}\natexlab{}.
\newblock \showarticletitle{Climate change policy networks: Why and how to
  compare them across countries}.
\newblock \bibinfo{journal}{\emph{Energy Research \& Social Science}}
  \bibinfo{volume}{45} (\bibinfo{year}{2018}), \bibinfo{pages}{258--265}.
\newblock


\bibitem[Zhao and Rosson(2009)]%
        {zhao2009HowWhyPeople}
\bibfield{author}{\bibinfo{person}{Dejin Zhao} {and} \bibinfo{person}{Mary~Beth
  Rosson}.} \bibinfo{year}{2009}\natexlab{}.
\newblock \showarticletitle{How and Why People {{Twitter}}: The Role That
  Micro-Blogging Plays in Informal Communication at Work}. In
  \bibinfo{booktitle}{\emph{Proceedings of the 2009 {{ACM International
  Conference}} on {{Supporting Group Work}}}} \emph{(\bibinfo{series}{{{GROUP}}
  '09})}. \bibinfo{publisher}{{Association for Computing Machinery}},
  \bibinfo{address}{{New York, NY, USA}}, \bibinfo{pages}{243--252}.
\newblock
\showISBNx{978-1-60558-500-0}
\urldef\tempurl%
\url{https://doi.org/10.1145/1531674.1531710}
\showDOI{\tempurl}


\end{thebibliography}

\appendix
\section{Additional Results}
\begin{figure}[h]
    \centering
    \includegraphics[width=0.9\textwidth]{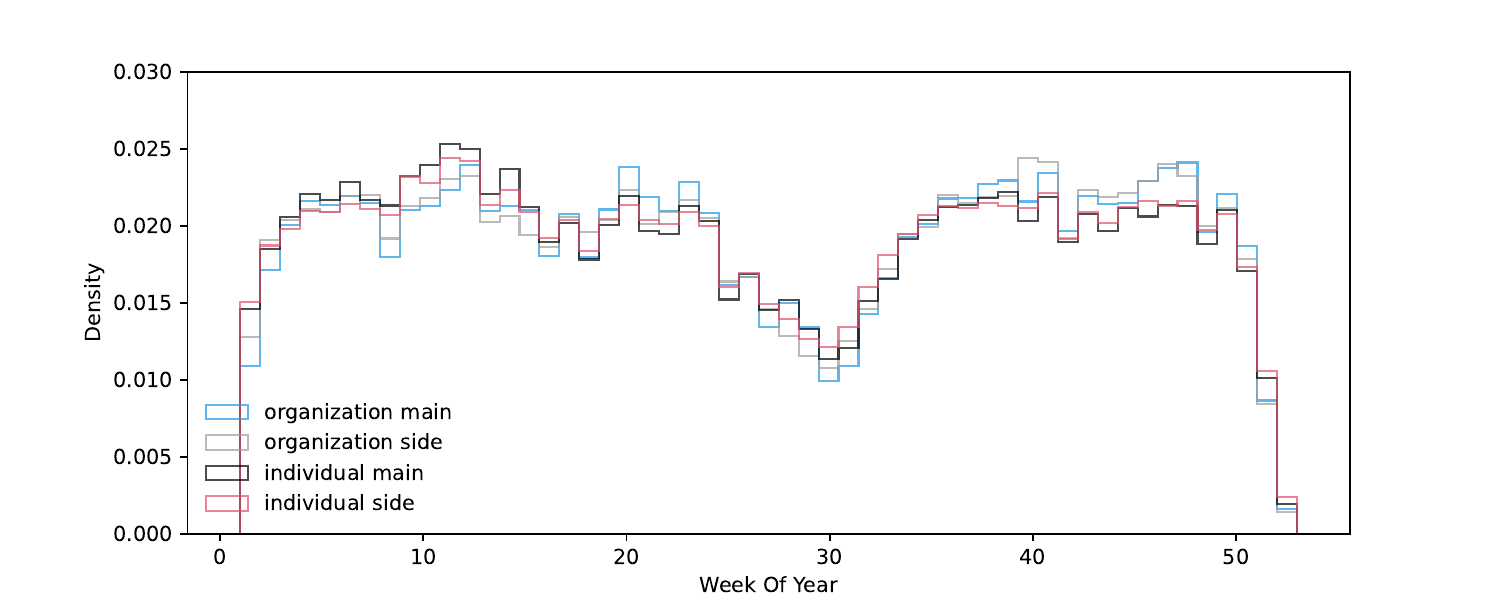}
    \caption{Activity Level by Week of Year}
    \label{fig:activity_level_org_main_by_hour_of_week}
\end{figure}

\subsection{Unweighted overlap}
\label{sec:woverlap}

For unweighted overlap, we measure how many edges two vertices have in common out of all the edges that they each have separately. Here, the edges could be to accounts outside of the organization. The unweighted overlap for vertices $i$ and $j$ is:
\begin{equation}
    O_{ij} = \frac{n_{ij}}{ (k_i - 1) + (k_j - 1) - n_{ij}}\,,     
\end{equation}
where $n_{ij}$ is the number of common edges between {i} and {j} and $k_i$ and $k_j$ are the total number of edges for $i$ and $j$, respectively. 

\begin{figure}[h]
    \centering
        \includegraphics[width=0.5\textwidth]{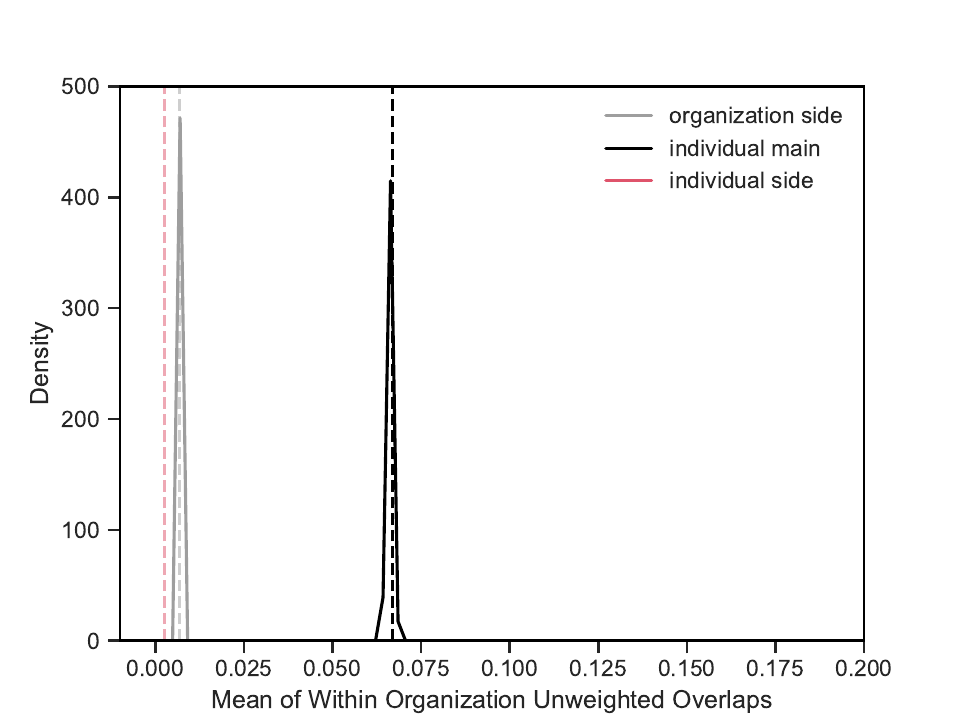}
    \caption{Unweighted Overlaps}
    \label{fig:unweighted_overlaps}
\end{figure}

Figure~\ref{fig:unweighted_overlaps} shows unweighted overalap results that are equivalent to weighted overlap in Figure~\ref{fig:densities_and_overlaps}.

\end{document}